\documentclass[10pt]{article}

\usepackage{amsmath}
\usepackage{amssymb}

\usepackage[pdftex]{graphicx}

\usepackage{cite}

\usepackage{color}
\usepackage{hyperref}

\usepackage[labelfont=bf,labelsep=period,justification=raggedright]{caption}

\makeatletter
\renewcommand{\@biblabel}[1]{\quad#1.}
\makeatother

\newcommand{\beginsupplement}{%
        \setcounter{table}{0}
        \renewcommand{\thetable}{S\arabic{table}}%
        \setcounter{figure}{0}
        \renewcommand{\thefigure}{S\arabic{figure}}%
     }
\date{}

\begin{document}

\begin{flushleft}
{\Large
\textbf{Deep Neural Networks Rival the Representation of Primate IT Cortex for Core Visual Object Recognition}
}
\\
Charles F.~Cadieu$^{1,\ast}$,
Ha Hong$^{1,2}$,
Daniel L. K. Yamins$^1$,
Nicolas Pinto$^1$,
Diego Ardila$^1$,
Ethan A. Solomon$^1$,
Najib J. Majaj$^1$,
James J. DiCarlo$^1$
\\
\bf{1} Department of Brain and Cognitive Sciences and McGovern Institute for Brain Research, Massachusetts Institute of Technology, Cambridge, MA 02139
\\
\bf{2} Harvard--MIT Division of Health Sciences and Technology, Institute for Medical Engineering and Science, Massachusetts Institute of Technology, Cambridge, MA 02139
\\
$\ast$ E-mail: Corresponding cadieu@mit.edu
\end{flushleft}

\section*{Abstract}
The primate visual system achieves remarkable visual object recognition performance even in brief presentations and under changes to object exemplar, geometric transformations, and background variation (a.k.a. core visual object recognition).  This remarkable performance is mediated by the representation formed in inferior temporal (IT) cortex.  In parallel, recent advances in machine learning have led to ever higher performing models of object recognition using artificial deep neural networks (DNNs).  It remains unclear, however, whether the representational performance of DNNs rivals that of the brain.  To accurately produce such a comparison, a major difficulty has been a unifying metric that accounts for experimental limitations such as the amount of noise, the number of neural recording sites, and the number trials, and computational limitations such as the complexity of the decoding classifier and the number of classifier training examples.  In this work we perform a direct comparison that corrects for these experimental limitations and computational considerations.  As part of our methodology, we propose an extension of ``kernel analysis'' that measures the generalization accuracy as a function of representational complexity.  Our evaluations show that, unlike previous bio-inspired models, the latest DNNs rival the representational performance of IT cortex on this visual object recognition task.  Furthermore, we show that models that perform well on measures of representational performance also perform well on measures of representational similarity to IT and on measures of predicting individual IT multi-unit responses.  Whether these DNNs rely on computational mechanisms similar to the primate visual system is yet to be determined, but, unlike all previous bio-inspired models, that possibility cannot be ruled out merely on representational performance grounds.

\section*{Author Summary}
Primates are remarkable at determining the category of a visually presented object even in brief presentations and under changes to object exemplar, position, pose, scale, and background.  To date, this behavior has been unmatched by artificial computational systems.  However, the field of machine learning has made great strides in producing artificial deep neural network systems that perform highly on object recognition benchmarks.  In this study, we measured the responses of neural populations in inferior temporal  (IT) cortex across thousands of images and compared the performance of neural features to features derived from the latest deep neural networks.  Remarkably, we found that the latest artificial deep neural networks achieve performance equal to the performance of IT cortex,  yielding a representational space in which images with objects of the same category are close, and objects of different categories are far apart even in the presence of large variations in object exemplar, position, pose, scale, and background.  We directly measured the object recognition abilities of high-level visual cortex in macaque monkey and compared its performance to the latest deep neural networks using measures of object recognition representational performance.  Furthermore, we show that the top-level features in these models exceed previous models in predicting the IT neural responses themselves.  This result indicates that the latest deep neural networks may provide insight into understanding primate visual processing.

\section*{Introduction}
Primate vision achieves a remarkable proficiency in object recognition, even in brief visual presentations and under changes to object exemplar, geometric transformations, and background variation.  Humans~\cite{Thorpe:1996wk} and macaques~\cite{FabreThorpe:1998te} are known to solve this task with high accuracy at low latency for presentation times shorter than 100 ms~\cite{Keysers:2001ff,Potter:2013go}.  This ability is likely related to the presence and rate of saccadic eye movements, which for natural viewing typically occur at a rate of one saccade every 200-250 ms~\cite{Andrews19992947}.  Therefore, when engaged in natural viewing the primate visual system is proficient at recognizing and making rapid and accurate judgements about the objects present within a single saccadic fixation.  While not encompassing all of primate visual abilities, this ability is an important subproblem that we operationally define and refer to as ``core visual object recognition''~\cite{DiCarlo:2012em}.

A key to this primate visual object recognition ability is the representation that the cortical ventral stream creates from visual signals from the eye.  The ventral stream is a series of cortical visual areas extending from primary visual area V1, through visual areas V2 and V4, and culminating in inferior temporal (IT) cortex.  At the end of the ventral stream, IT cortex creates a representation of visual stimuli that is selective for object identity and tolerant to nuisance parameters such as object position, scale, pose, and background~\cite{Desimone:1984wk,kobatake1994neuronal,hung2005fast,Rust:2010uk}.  The responses of IT neurons are remarkable because they indicate that the ventral stream has transformed the complicated, non-linear object recognition problem at the retinae into a new neural representation that separates objects based on their category~\cite{dicarlo2007untangling,DiCarlo:2012em}.  Results using linear classifiers have shown that the IT neural representation creates a simpler object recognition problem that can often be solved with a linear function predictive of object category~\cite{hung2005fast,Rust:2010uk}.  It is thought that this transformation is achieved through the ventral stream by a series of recapitulated modules that each produce a non-linear transformation of their input that becomes selective for objects and tolerant to nuisance variables unrelated to object identity~\cite{DiCarlo:2012em}.

A number of bio-inspired models have sought to replicate the phenomenology observed in the primate ventral stream (see e.g. \cite{Fukushima:1980iz,Riesenhuber1999,Stringer:2002hv,serre2007,Pinto:2009ho}) and recent, related models in the machine learning community, generally referred to as ``deep neural networks'' share many properties with these bio-inspired models.  The computational concepts utilized in these models date back to early models of the primate visual system in the work of Hubel and Wiesel~\cite{DHHubel:1962vm,hubel1968receptive}, who hypothesized that within primary visual cortex more complex functional responses (``complex'' cells) were constructed from more simplistic responses (``simple'' cells).  Models of biological vision have extended this hypothesis by suggesting that higher visual areas recapitulate this mechanism and form a hierarchy~\cite{Fukushima:1980iz,Riesenhuber1999,Perrett:1993hl,Mel:1997cz,wallis1997,Serre:2007jy}.  In the last few years, a series of visual object recognition systems have been produced that utilize deep neural networks and have achieved state-of-the-art performance on computer vision benchmarks (see e.g. \cite{le2011building,Krizhevsky:2012wl,Zeiler:2013ws,Sermanet:2013vi}).  These deep neural networks implement architectures containing successive layers of operations that resemble the simple and complex cell hierarchy first described by Hubel and Wiesel.  However, unlike previous bio-inspired models, these latest deep neural networks contain many layers of computation (typically 7-9 layers, while previous models contained 3-4) and adapt the parameters of the layers using supervised learning on millions of object-labeled images (the parameters of previous models were either hand-tuned, adapted through unsupervised learning, or trained on just thousands of labeled images).  Given the increased complexity of these deep neural networks and the dramatic increases in performance over previous models, it is relevant to ask, ``how close are these models to achieving object recognition representational performance that is similar to that observed in IT cortex?''  In this work we seek to address this question.

Our methodology directly compares the representational performance of IT cortex to deep neural networks and overcomes the shortcoming of previous comparisons.  There are four areas where our approach has advantages over previous attempts.  Although previous attempts have addressed one or two of these shortcomings, none has addressed all four.  First, previous attempts have not corrected for a number of experimental limitations including the amount of experimental noise, the number of recorded neural sites, or the number of recorded stimulus presentations (see e.g.~\cite{hung2005fast,Rust:2010uk,Yamins:2014gi}).  Our methodology makes explicit these limitations by either correcting for, or modifying model representations to arrive at a fair comparison to neural representation.  We find that these corrections have a dramatic effect on our results and shed light on previous comparisons that we believe may have been misleading.

Second, previous attempts have utilized fixed complexity classifiers and have not addressed the relationship between classifier complexity and decision boundary accuracy (see e.g.~\cite{hung2005fast,Rust:2010uk,Yamins:2014gi}).  In our methodology we utilize a novel extension of ``kernel analysis,'' formulated in the works of~\cite{Braun:2006va,Braun:2008ul,Montavon:2011wp}, to measure the accuracy of a representation as a function of the complexity of the task decision boundary.  This allows us to identify representations that achieve high accuracy for a given complexity and avoids a measurement confound that arises when using cross-validated accuracy: the decision boundary's complexity and/or constraints are dependent on the size and choice of the training dataset, factors that can strongly affect accuracy scores.  

Third, previous attempts have not measured the variations in the neural or model spaces that are relevant to class-level object classification~\cite{Kriegeskorte:2008vz}.  For example the work in~\cite{Kriegeskorte:2008vz} examined the variation present in neural populations to visual stimuli presentations and compared this variation to the variation produced in model feature spaces to the same stimuli.  This methodology does not address representational performance and does not provide an accuracy-complexity analysis (however, see \cite{NikolausKriegeskorte:2008bz} and \cite{Mur:2012vq}, for discussion of methodologies to account for dissimilarity matrices by class-distance matrices).  Our methodology of analyzing absolute representational performance using kernel analysis provides a novel and complementary finding to the results in \cite{NikolausKriegeskorte:2008bz,Yamins:2013vu,Yamins:2014gi}.  Because of this complementarity, in this paper we also directly measure the amount of IT neural variance captured by deep neural networks as IT encoding models and by measuring representational similarity.

Finally, our approach utilizes a dataset that is an order of magnitude larger than previous datasets, and captures a degree of stimulus complexity that is critical for assessing IT representational performance.  For example, the analysis in~\cite{Rust:2010uk} utilized 150 images and the comparison in~\cite{Kriegeskorte:2008vz} utilized 96 images, while in this work we utilize an image set of 1960 images.  The larger number of images allows our dataset to span and sample a relatively high degree of stimulus variation, which includes variation due to object exemplar, geometric transformations (position, scale, and rotation/pose) and background.  \emph{Importantly this variation is critical to distinguish between models based on object classification performance}: only in the presence of high variation are models distinguishable from each other~\cite{Pinto:2008gj,Pinto:2011iw} and from IT~\cite{Yamins:2014gi}.

In this work, we propose an object categorization task and establish measurements of human performance for brief visual presentations.  We then present our novel extension of kernel analysis and show that the latest deep neural networks achieve higher representational performance on this visual task compared to previous generation bio-inspired models.  We next compare model representational performance to the IT cortex neural representation on the same task and images by matching the number of model features to the number of IT recordings and to the amount of observed experimental noise for both multi-unit recordings and single-unit recordings.  We find that the latest DNNs match IT performance whereas previous models significantly lag the IT neural representation.  In addition, we replicate the findings using a linear classifier approach.  Finally, we show that the latest DNNs also provide compelling models of the actual IT neural response by measuring encoding model predictions and performing a representational similarity analysis.  We conclude with a discussion of the limitations of the current approach and future directions for studying models of visual recognition and primate object recognition.

\section*{Results}
To evaluate the question of representational performance we must first make a choice about the task to be analyzed.  The task we examine here is visual object category recognition in a natural duration fixation.  This task is a well studied subproblem in visual perception and tests a core problem of visual perception: context independent basic-level object recognition within brief visual presentation.  The task is to determine the category of an object instance that is presented under the effect of image variations due to object exemplar, geometric transformations (position, scale, and rotation/pose), and background.  This task is well supported by behavioral measurements: humans~\cite{Thorpe:1996wk} and macaques~\cite{FabreThorpe:1998te} are known to solve this task with high proficiency.  It is well supported by neural measurements: evidence from IT cortex indicates that the neural representation supports and performs highly on this task~\cite{Weiskrantz:1984ef}.  Furthermore, this task provides a computationally challenging problem on which previous computational models have been shown to severely underperform~\cite{Pinto:2008gj,Pinto:2011iw}.  Therefore, this task is difficult computationally and is performed at high proficiency by primates, with evidence that the primate ventral visual stream produces an effective representation in IT cortex.  
\begin{figure*}[t]
\begin{center}
\includegraphics[width=\linewidth]{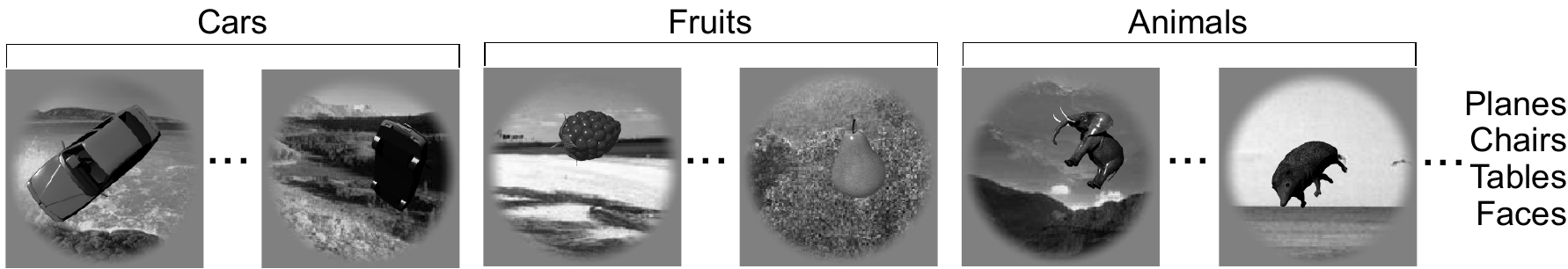}
\end{center}
\caption{{\bf Example images used to measure object category recognition performance.}  Two of the 1960 tested images are shown from the categories Cars, Fruits, and Animals (we also tested the categories Planes, Chairs, Tables, and Faces).  Variability within each category consisted of changes to object exemplar (e.g. 7 different types of Animals), geometric transformations due to position, scale, and rotation/pose, and changes to background (each background image is unique).}
\label{fig:images}
\end{figure*}

Methodologically, the task is defined through an image generation process.  An image is constructed by first choosing one of seven categories, then one of seven 3D object exemplars from that category, then a randomly chosen background image (each background image is used only once), and finally the variation parameters are drawn from a distribution to span two full octaves of scale variation, the full width of the image for translation variation, and the full sphere for pose variation.  For each object exemplar we generated 40 unique images using this process, resulting in 1960 images in total.  See Figure \ref{fig:images} for example images organized by object category and Methods for further description of the image generation process.  The resulting image set has several advantages and disadvantages.  Advantageously, this procedure eliminates dependencies between objects and backgrounds that may be found in real-world images~\cite{Oliva:2007ui}, and introduces a controlled amount of variability or difficulty in the task, which we have used to produce image datasets that are known to be difficult for algorithms~\cite{Pinto:2008gj,Pinto:2010hvm,Pinto:2011iw}.  Though arguably not fully ``natural'', the resulting images are highly complex (see Discussion for further advantages and disadvantages).

In evaluating the neural representational performance we must also define the behavioral context within which the neural representation supports behavior.  This definition is important because it determines specific choices in the experimental setup.  The behavioral context that we seek to address is a sub-problem of general visual behavior: vision in a natural duration fixation, or visual object recognition within one fixation without contextual influence, eye movements, or shifts in attention (also called ``core visual object recognition''~\cite{DiCarlo:2012em}).  In our neural experiments we have chosen a presentation time of 100 milliseconds (ms) so as to be relevant for this behavior (see Discussion for further justification and Supporting Information (SI) for behavioral measurements on this task).

As a first step to evaluate the neural representation, we recorded multi-unit and single-unit neural activity from awake behaving rhesus macaques during passive fixation.  We recorded activity using large scale multi-electrode arrays placed in either IT cortex or visual area V4.  To create a neural feature vector, which we use to assess object representational performance, we presented each image (1960 images in total) for 100 ms and measured the normalized, background subtracted firing-rate in a window from 70 ms to 170 ms post image onset, averaged over 47 repetitions (see Methods).  Over two macaques we measured 168 multi-unit sites in IT cortex, and 128 multi-unit sites in V4.  From these recordings we also isolated single-units from IT and V4 cortex.  Using conservative criteria (see Methods), we isolated 40 single-units from IT and 40 single-units from V4 with 6 repetitions per image for each single-unit.

To evaluate the performance of neural or model representations we utilize a novel extension of kernel analysis.  Kernel analysis evaluates the efficacy of the representation by measuring how the precision of the category regression problem changes as we allow the complexity of the regression function to increase~\cite{Montavon:2011wp}.  Intuitively, more effective representations will achieve higher precision at the same level of complexity because they have removed irrelevant variability from the original representational space (here irrelevant variability in the original space is due to object exemplar, geometric transformation, and background).  To measure precision vs. complexity of the regression function, we perform kernel ridge regression using a Gaussian kernel (see Methods for details).  We define complexity as the inverse of the regularization parameter ($1/\lambda$) and precision as 1 minus the normalized mean-squared leave-one-example-out generalization error, such that a precision value of 0 is chance performance and 1 is perfect performance.  The regularization parameter restricts the complexity of the resulting regression function.  By choosing a Gaussian kernel we can move between regression functions that are effectively linear, to functions that interpolate between the data points\footnote{Complex regression functions may not generalize if there are not enough training examples (known as ``sample complexity''), which will result in saturation or reduction in accuracy as complexity increases.} (a ``complex'' regression function)\cite{Keerthi:2003uw}.

\begin{figure}
\begin{center}
\includegraphics[width=.56\linewidth]{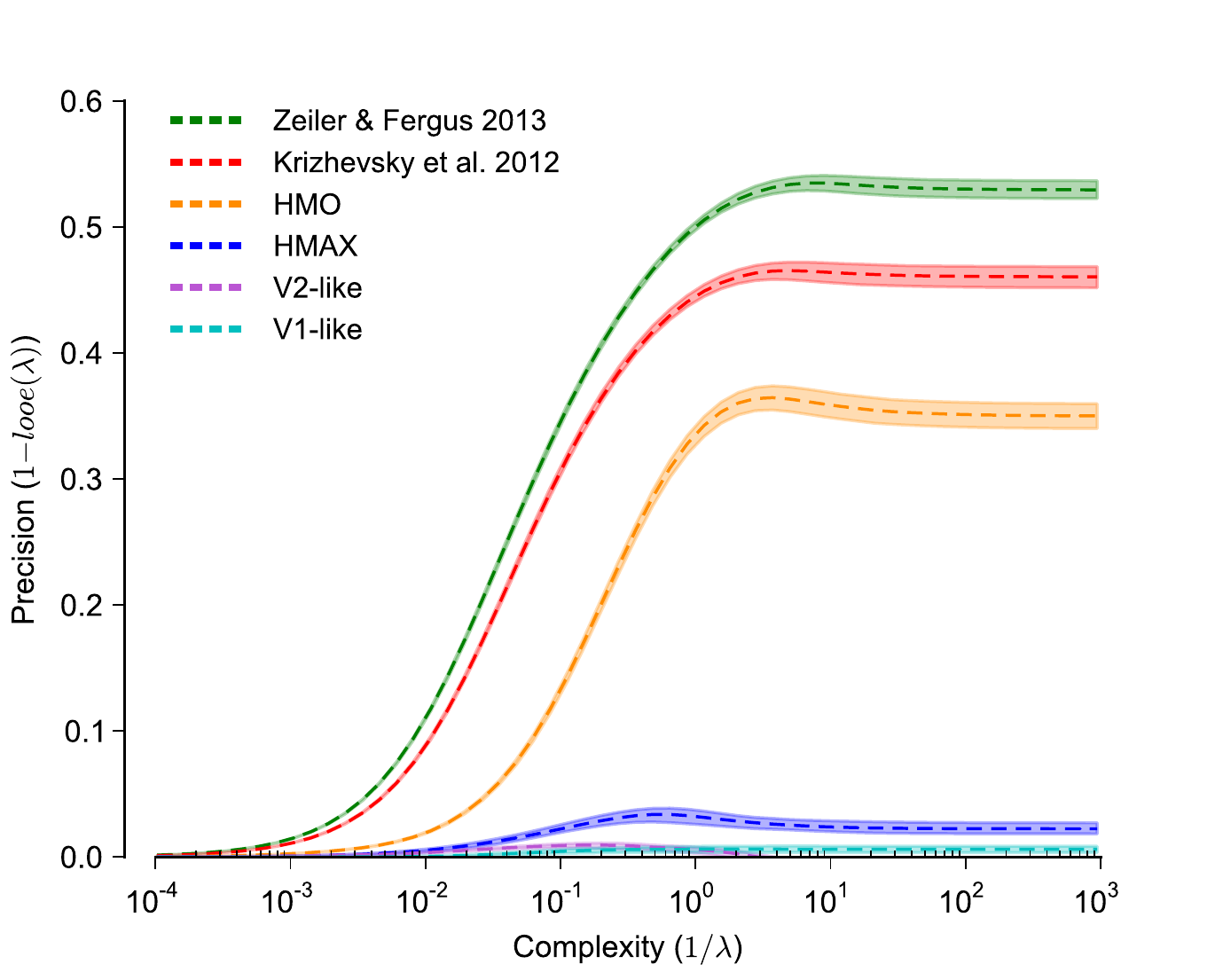}
\end{center}
\caption{{\bf Kernel analysis curves of model representations.}  Precision, one minus loss ($1 - looe(\lambda)$), is plotted against complexity, the inverse of the regularization parameter ($1/\lambda$).  Shaded regions indicate the standard deviation of the measurement over image set randomizations, which are often smaller than the line thickness.  The Zeiler \& Fergus 2013, Krizhevsky et al. 2012 and HMO models are all hierarchical deep neural networks.  HMAX~\cite{Mutch:2008eo} is a model of the ventral visual stream and the V1-like~\cite{Pinto:2008gj} and V2-like~\cite{Freeman:2011gl} models attempt to replicate response properties of visual areas V1 and V2, respectively.  These analyses indicate that the task we are measuring proves difficult for V1-like and V2-like models, with these models barely moving from 0.0 precision for all levels of complexity.  Furthermore, the HMAX model, which has previously been shown to perform relatively well on object recognition tasks, performs only marginally better.  Each of the remaining deep neural network models performs drastically better, with the Zeiler \& Fergus 2013 model performing best for all levels of complexity.  These results indicate that the visual object recognition task we evaluate is computationally challenging for all but the latest deep neural networks.}
\label{fig:ka_models}
\end{figure}
We compared the neural representation to three convolutional DNNs and three other biologically relevant representations.  Note that the development of these representations did not utilize the 1960 images we use here for testing in any way.  The three recent convolutional DNNs we examine are described in Krizhevsky et al. 2012~\cite{Krizhevsky:2012wl}, Zeiler \& Fergus 2013~\cite{Zeiler:2013ws}, and Yamins et al. 2014~\cite{Yamins:2013vu,Yamins:2014gi}.  The Krizhevsky et al. 2012 and Zeiler \& Fergus 2013 DNNs are of note because they have each successively surpassed the state-of-the-art performance on the ImageNet Large Scale Visual Recognition Challenge (ILSVRC) datasets.\footnote{Note that results have continued to improve on this challenge since we ran our analysis.  See \url{http://www.image-net.org/} for the latest results.}  The DNN presented in Yamins et al. 2014~\cite{Yamins:2014gi} is created using a supervised optimization procedure called hierarchical modular optimization (we refer to this model by the abbreviation HMO).  The HMO DNN has been shown to match closely representational dissimilarity matrices of the ventral stream and to be predictive of IT and V4 neural responses~\cite{Yamins:2014gi}.  We also evaluated an instantiation of the HMAX model of invariant object recognition that uses sparse localized features~\cite{Mutch:2008eo} and has previously been shown to be a relatively high performing model among artificial systems \cite{Pinto:2009ho}.  Finally, we also evaluated a V2-like model and a V1-like model that each attempt to capture a first-order account of secondary (V2)~\cite{Freeman:2011gl} and primary visual cortex (V1)~\cite{Pinto:2008gj}, respectively.

Each of the three convolutional DNNs was developed, implemented, and trained by their respective researchers and for those developed outside of our group we obtained features from each DNN computed on our test images.  The convolutional DNN described in Krizhevsky et al. 2012 \cite{Krizhevsky:2012wl} was trained by supervised learning on the ImageNet 2011 Fall release ($\sim$15M images, 22K categories) with additional training on the LSVRC-2012 dataset (1000 categories).  The authors computed the features in the penultimate layer of their model (4096 features) on the 1960 images we used to measure the neural representation.  The similar 8-layer deep neural network of Zeiler \& Fergus 2013 \cite{Zeiler:2013ws} was trained using supervised learning on the LSVRC-2012 dataset augmented with random crops and left-right flips.  This model took advantage of hyper-parameter tuning informed by visualizations of the intermediate network layers.  The 4096 dimensional feature representation was produced by taking the penultimate layer features and averaging them over 10 image crops (the 4 corners, center, and horizontal flips for each).  The model of Yamins et al. 2014~\cite{Yamins:2014gi} is an extension of the high-throughput optimization strategy described in~\cite{Pinto:2009ho} that produces a heterogeneous combination of hierarchical convolutional models optimized on a supervised object recognition task through hyperparameter optimization using boosting and error-based reweighing (see \cite{Yamins:2014gi} for details).  The total output feature space per image for the HMO model is 1250 dimensional.

\begin{figure}
\begin{center}
\includegraphics[width=\linewidth]{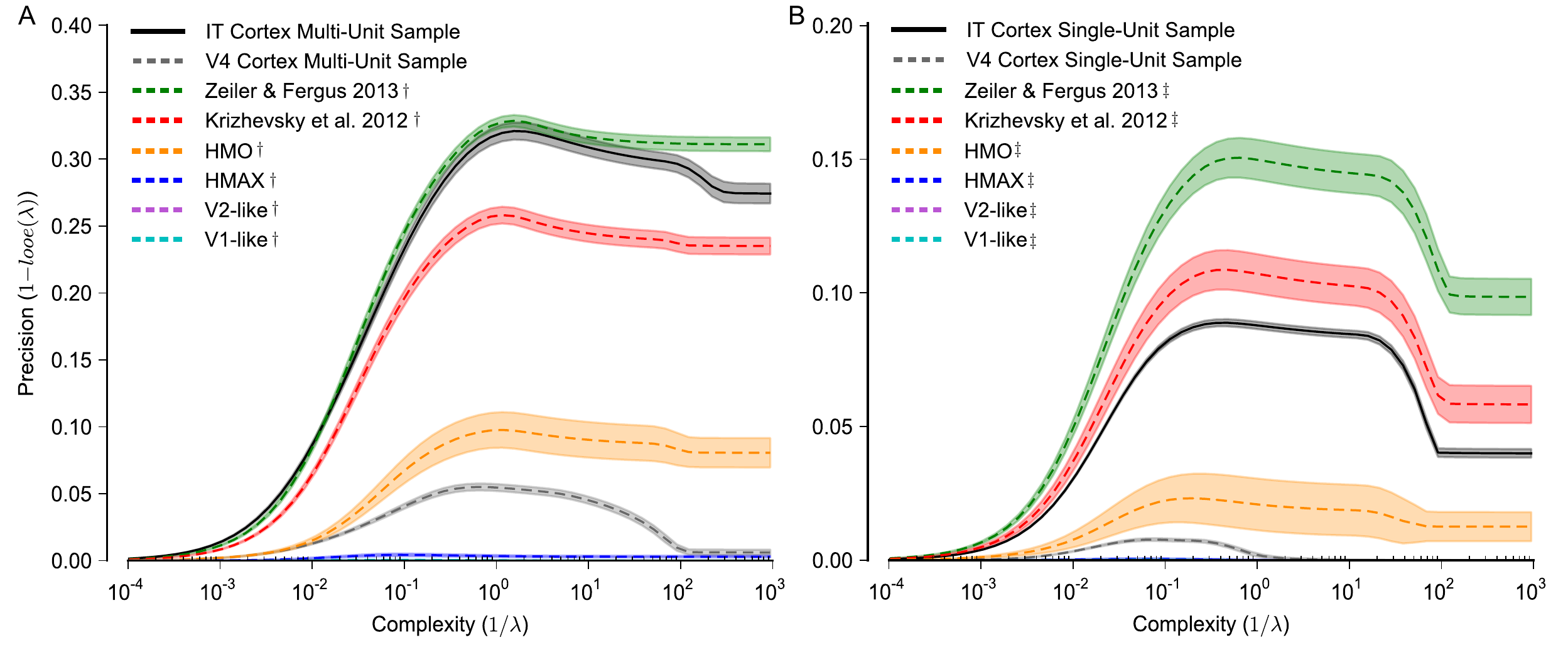}
\end{center}
\caption{{\bf Kernel analysis curves of sample and noise matched neural and model representations.}  Plotting conventions are the same as in Figure \ref{fig:ka_models}.  Multi-unit analysis is presented in panel A and single-unit analysis in B.  Note that the model representations have been modified such that they are both subsampled and noisy versions of those analyzed in Figure \ref{fig:ka_models} and this modification is indicated by the $\dagger$ symbol for noise matched to the multi-unit IT cortex sample and by the $\ddagger$ symbol for noise matched to the single-unit IT cortex sample.  To correct for sampling bias, the multi-unit analysis uses 80 samples, either 80 neural multi-units from V4 or IT cortex, or 80 features from the model representations, and the single-unit analysis uses 40 samples.  To correct for experimental and intrinsic neural noise, we added noise to the subsampled model representation (no additional noise is added to the neural representations) that is commensurate to the observed noise from the IT measurements.  Note that we observed similar noise between the V4 and IT Cortex samples and we do not attempt to correct the V4 cortex sample of the noise observed in the IT cortex sample.  We observed substantially higher noise levels in IT single-unit recordings than multi-unit recordings due to both higher trial-to-trial variability and more trials for the multi-unit recordings.  All model representations suffer decreases in accuracy after correcting for sampling and adding noise (compare absolute precision values to Figure \ref{fig:ka_models}).  All three deep neural networks perform significantly better than the V4 cortex sample.  For the multi-unit analysis (A), IT cortex sample achieves high precision and is only matched in performance by the Zeiler \& Fergus 2013 representation.  For the single-unit analysis (B), both the Krizhevsky et al. 2012 and the Zeiler \& Fergus 2013 representations surpass the IT representational performance.}
\label{fig:ka_neural}
\end{figure}
Before comparing the representational performance of the neural and model representations, we first evaluate the absolute representational performance of these models on the task to verify that the task we have chosen is computationally difficult.  As described in our previous work~\cite{Pinto:2008gj}, we determined that a task is computationally difficult if ``simple'' computational models fail on the task.  For the models tested here, the V1-like and V2-like models represent these computationally simple models.  Using kernel analysis we evaluated both the DNNs and the bio-inspired models on the task and plot the precision vs. the complexity curves for each model representation in Figure \ref{fig:ka_models}.  This analysis indicates that both the V1-like and V2-like models perform near chance on this task over the entire range of complexity.  Furthermore, the HMAX model performs only slightly better on this task.  If we reduce the difficulty of the task by reducing the magnitude range of the variations we introduce (not shown here, but see \cite{Pinto:2008gj,Cadieu:2013wa} for such an analysis), these models are known to perform well on this task; therefore, it is object recognition under variation that makes this problem difficult, and the magnitude range we have chosen for this task is quite difficult in that the HMAX model performs poorly.  In contrast, the three DNNs perform at much higher precision levels over the complexity range.  A clear ranking is observed with the Zeiler \& Fergus 2013~\cite{Zeiler:2013ws} model followed by the Krizhevsky et al. 2012~\cite{Krizhevsky:2012wl} model and the HMO model~\cite{Yamins:2014gi}.  These results indicate that these models outperform models of early visual areas, and we next ask which model, if any, can match the performance of high-level visual area IT.

\begin{figure}[t]
\begin{center}
\includegraphics[width=\linewidth]{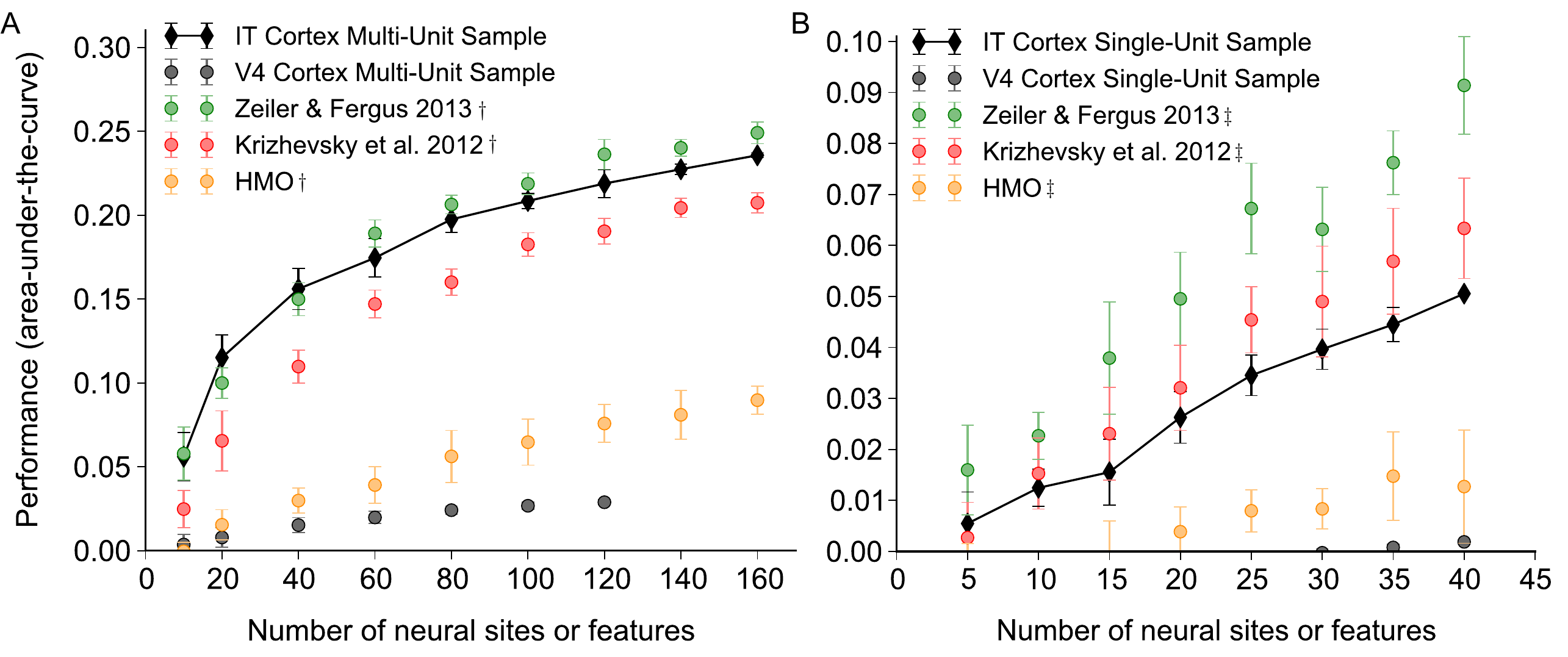}
\end{center}
\caption{{\bf Effect of sampling the neural and noise-corrected model representations.}  We measure the area-under-the-curve of the kernel analysis measurement as we change the number of neural sites (for neural representations), or the number of features (for model representations).  Measured samples are indicated by filled symbols and measured standard deviations indicated by error bars.  Multi-unit analysis is shown in panel A and single-unit analysis in B.  The model representations are noise corrected by adding noise that is matched to the IT multi-unit measurements (A, as indicated by the $\dagger$ symbol) or single-unit measurements (B, as indicated by the $\ddagger$ symbol).  For the multi-unit analysis, the Zeiler \& Fergus 2013 representation rivals the IT cortex representation over our measured sample.  For the single-unit analysis, the Krizhevsky et al. 2012 representation rivals the IT cortex representation for low number of features and slightly surpasses it for higher number of features.  The Zeiler \& Fergus 2013 representation surpasses the IT cortex representation over our measured sample.}
\label{fig:subsampled}
\end{figure}

In order to directly compare the representational performance of the IT neural representation to the model representations we take a number of steps to produce a fair comparison.  The experimental procedure that we used to measure the neural representation is limited by the number of neural samples (sites or number of neurons) that we can measure and by noise induced by uncontrolled experimental variability and/or intrinsic neural noise.  To equalize the sampling between the neural representation and the model representations we fix the number of neural samples (80 for the multi-unit analysis and 40 for the single-unit analysis) and model features (we will vary this number in later experiments, Figure \ref{fig:subsampled}).  To correct for the observed experimental noise, we add noise to the model representations.  To add noise to the models we estimate an experimental neural noise model.  Following the observation that spike counts of neurons are approximately Poisson~\cite{Tolhurst:1983wa,Shadlen:1998ta} and similar analyses of our own recordings, we model response variability as being proportional to the mean response.  Precisely, the estimated noise model is additive to the mean response and is a zero-mean Gaussian random variable with variance being a linear function of the mean response.  We estimate the parameters of the noise model from the empirical distribution of multi-unit responses and single-unit responses.  Note that our empirical estimate of these quantities is influenced by both uncontrolled experimental variability (e.g. variability across recording sessions) as well as intrinsic neural noise.  See Methods for further description and Figure~\ref{fig:noisetest} for a verification that the noise model reduces performance more greatly than empirical noise, thus demonstrating that the noise model is conservative and over-penalizes models.  To produce noise-matched model representations, we sample the model response dependent noise and measure the representational performance of the resulting representation using kernel analysis.  We repeat this procedure 10 times to measure the variability produced by the additive noise model.

We compare the sample and noise corrected model representations to the multi-unit neural representations in Figure \ref{fig:ka_neural}A.  The kernel analysis curves are plotted for neural and model representations sampled at 80 neural samples or 80 model features, respectively.  The model representations have been corrected for the neural noise observed in the multi-unit IT neural measurement.  Note that we do not attempt to correct the V4 sample to the noise level observed in IT because we observed similar noise between the V4 and IT neural measurements and each sample is averaged over the same number of trials (47 trials).  Compared to the model representational performance in Figure \ref{fig:ka_models}, model performance is reduced because of the subsampling and because of the added noise correction (without added noise and subsampling maximum precision is above 0.5 and with noise and subsampling does not pass 0.35).  Consistent with previous work~\cite{Rust:2010uk,Pinto:2011iw}, we observed that the sampled IT neural representation significantly exceeds the similarly-sampled V4 neural representation.  Unsurprisingly, HMAX, V2-like, and V1-like representations perform near chance.  All three recent DNNs perform better than the V4 representation.  The IT representation performs quite well, especially considering the sampling and noise limitations of our recordings and would be quite competitive if directly compared to the model results in Figure \ref{fig:ka_models}.  After correcting for sampling and noise, the IT representation is only matched by the top performing DNN of Zeiler \& Fergus 2013.  Interestingly, this relationship holds for the entire complexity range.

We present the equivalent representational comparison between models and neural representations for the single-unit neural recordings in Figure \ref{fig:ka_neural}B.  Because of the increased noise and fewer trials collected for the single-unit measurements compared to our multi-unit measurements, the single-unit noise and sample corrected model representations achieve lower precision vs. complexity curves than under the multi-unit noise and sample correction (compare to Figure \ref{fig:ka_neural}A).  This analysis shows that the single-unit IT representation performs better than the HMO representation, slightly worse than the Krizhevsky et al. 2012 representation, and is outperformed by the Zeiler \& Fergus 2013 \cite{Zeiler:2013ws} representation.

In Figures ~\ref{fig:subsampled}A and~\ref{fig:subsampled}B we analyze the representational performance as a function of neural sites or model features for multi-unit and single-unit neural measurements.  To achieve a summary number from the kernel analysis curves we compute the area-under-the-curve and we omit the HMAX, V2-like, and V1-like models because they are near zero performance in this regime.  In Figure \ref{fig:subsampled}A we vary the number of multi-unit recording samples and the number of features.  Just as in Figure \ref{fig:ka_neural}A, we correct for neural noise by adding a matched neural noise level to the model representations.  Figure \ref{fig:subsampled}A indicates that the representational performance relationship we observed at 80 samples is robust between 10 samples and 160 samples.  Figure \ref{fig:subsampled}B indicates that the performance of the IT single-unit representation is comparatively worse than the multi-unit, with the single-unit representation falling below the performance of the Krizhevsky et al. 2012 representation for much of the range of our analysis.

These results indicate that after correcting for noise and sampling effects, the Zeiler \& Fergus 2013 DNN rivals the performance of the IT multi-unit representation and that both the Krizhevsky et al. 2012 and Zeiler \& Fergus 2013 DNNs surpasses the performance of the IT single-unit representation.  The performance of these two DNNs in the low-complexity regime is especially interesting because it indicates that they perform comparably to the IT representation in the low-sample regime (i.e. low number of training examples), where restricted representational complexity is essential for generalization (e.g. \cite{Evgeniou:2000fx}).

\begin{figure}[t]
\begin{center}
\includegraphics[width=.6\linewidth]{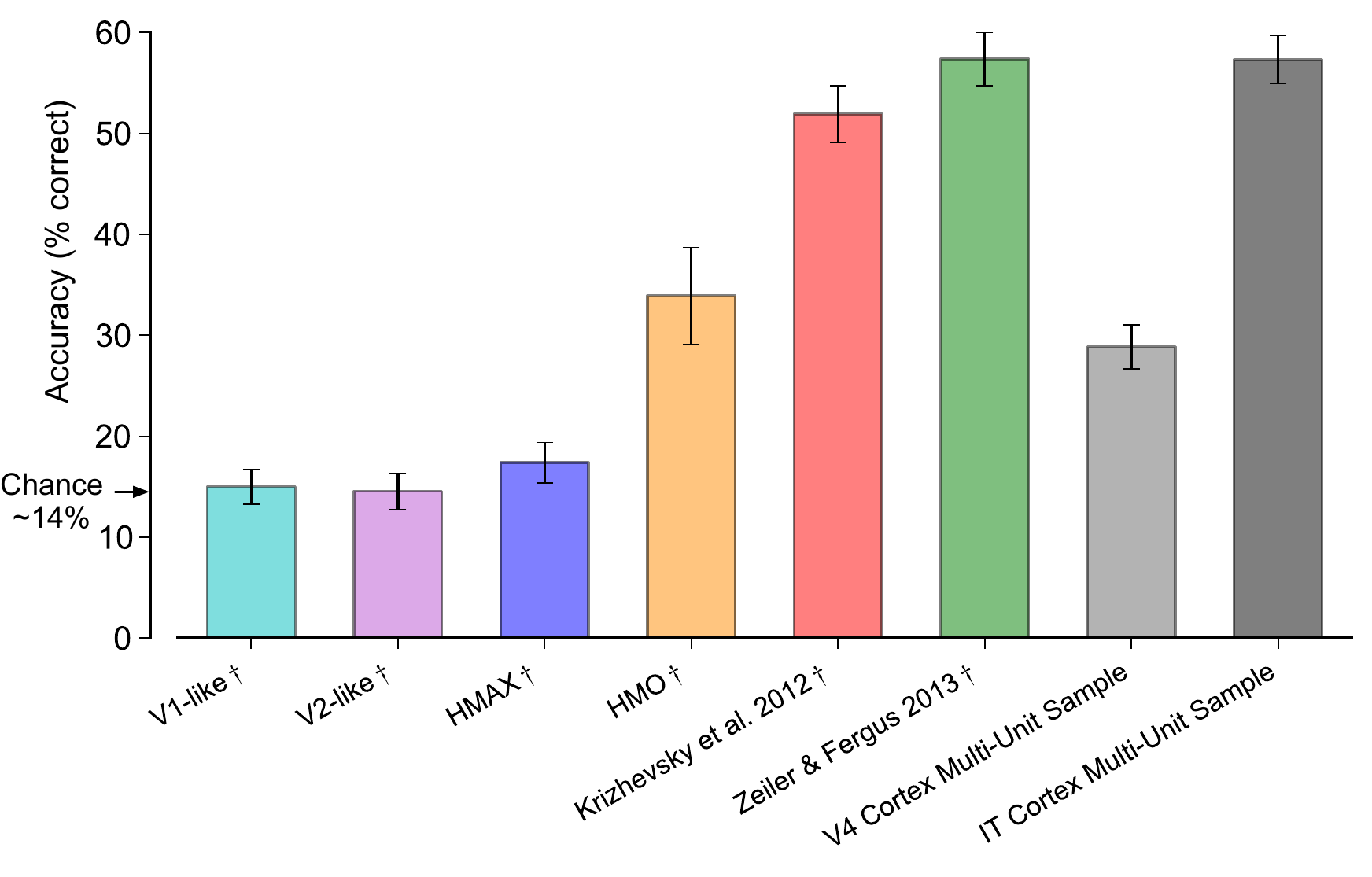}
\end{center}
\caption{{\bf Linear-SVM generalization performance of neural and model representations.}  Testing set classification accuracy averaged over 10 randomly-sampled test sets is plotted and error bars indicate standard deviation over the 10 random samples.  Chance performance is $\sim$14.3\%.  V4 and IT Cortex Multi-Unit Sample are the values measured directly from the neural samples.  Following the analysis in Figure~\ref{fig:ka_neural}A, the model representations have been modified such that they are both subsampled and have noise added that is matched to the observed IT multi-unit noise.  We indicate this modification by the $\dagger$ symbol.  Both model and neural representations are subsampled to 80 multi-unit samples or 80 features.  Mirroring the results using kernel analysis, the IT cortex multi-unit sample achieves high generalization accuracy and is only matched in performance by the Zeiler \& Fergus 2013 representation.}
\label{fig:svm}
\end{figure}

To verify the results of the kernel analysis procedure we measured linear-SVM generalization performance on the same task for each neural and model representation (Figure~\ref{fig:svm}).  We used a cross-validated procedure to train the linear-SVM on 80\% of the images and test on 20\% (regularization parameters were estimated from the training set).  We repeated the procedure for 10 randomizations of the training-testing split.  The linear-SVM results reveal a similar relationship to the results produced using kernel analysis (Figure~\ref{fig:ka_neural}A).  This indicates that the Zeiler \& Fergus 2013 representation achieves generalization comparable to the IT multi-unit neural sample for a simple linear decision boundary.  We also found near identical results to kernel analysis for the single-unit analyses and the analysis of performance as a function of the number of neural sites or features (see Figure~\ref{fig:svm_subsampled}).

\begin{figure}[t]
\begin{center}
\includegraphics[width=.95\linewidth]{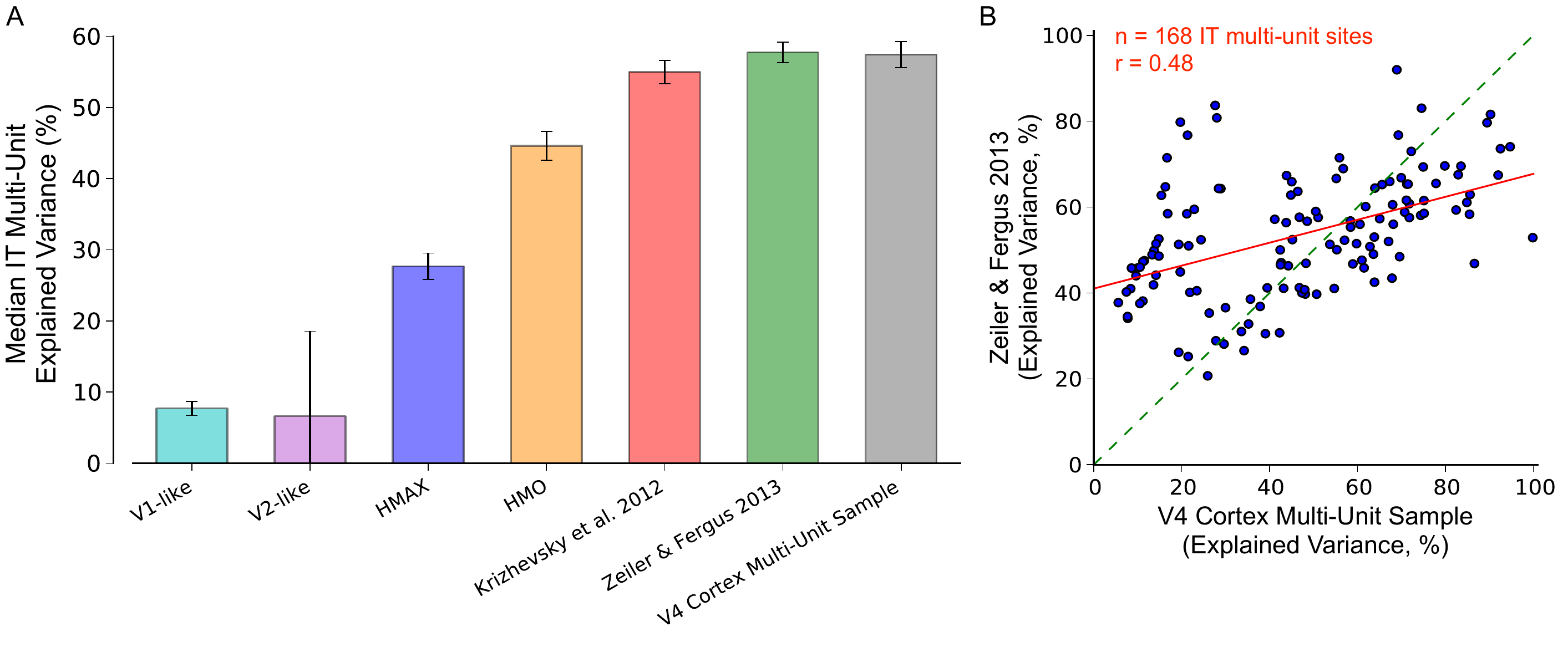}
\end{center}
\caption{{\bf Neural and model representation predictions of IT multi-unit responses.}  A) The median predictions of IT multi-unit responses averaged over 10 train/test splits is plotted for model representations and V4 multi-units.  Error bars indicate standard deviation over the 10 train/test splits.  Predictions are normalized to correct for trial-to-trial variability of the IT multi-unit recording and calculated as percentage of explained, explainable variance.  The HMO, Krizhevsky et al. 2012, and Zeiler \& Fergus 2013 representations achieve IT multi-unit predictions that are comparable to the predictions produced by the V4 multi-unit representation. B) The mean predictions over the 10 train/test splits for the V4 cortex multi-unit sample and the Zeiler \& Fergus 2013 DNN are plotted against each other for each IT multi-unit site.}
\label{fig:itfit}
\end{figure}

While the goal of our analysis has been to measure representational performance of neural and machine representations it is also informative to measure neural encoding metrics and measures of representational similarity.  Such analyses are complementary because representational performance relates to the task goals (in this case category labels) and encoding models and representational similarity metrics are informative about a model's ability to capture image-dependent neural variability, even if this variability is unrelated to task goals.  We measured the performance of the model representations as encoding models of the IT multi-unit responses by estimating linear regression models from the model representations to the IT multi-unit responses.  We estimated models on 80\% of the images and tested on 20\%, repeating the procedure 10 times (see Methods).  The median predictions averaged over the 10 splits are presented in Figure~\ref{fig:itfit}A.  For comparison, we also estimated regression models using the V4 multi-unit responses to predict IT multi-unit responses.  The results show that the Krizhevsky et al. 2012 and the Zeiler \& Fergus 2013 DNNs achieve higher prediction accuracies than the HMO model, which was previously shown to achieve high predictions on a similar test~\cite{Yamins:2014gi}.  These predictions are similar in explained variance to the predictions achieved by V4 multi-units.  However, no model is able to fully account for the explainable variance in the IT multi-unit responses.  In Figure~\ref{fig:itfit}B we show the mean explained variance of each IT multi-unit site as predicted by the V4 cortex multi-unit sample and the Zeiler \& Fergus 2013 DNN.  There is a relatively weak relationship between the encoding performance of the neural V4 and DNN representations ($r\!=\!0.48$ between V4 and Zeiler \& Fergus 2013, compared to $r\!=\!0.96$ and $r\!=\!0.74$ for correlations between Krizhevsky et al. 2012 and Zeiler \& Fergus 2013, and HMO and Zeiler \& Fergus 2013, respectively), indicating that V4 and DNN representations may account for different sources of variability in IT (see Discussion).

\begin{figure}
\begin{center}
\includegraphics[width=.6\linewidth]{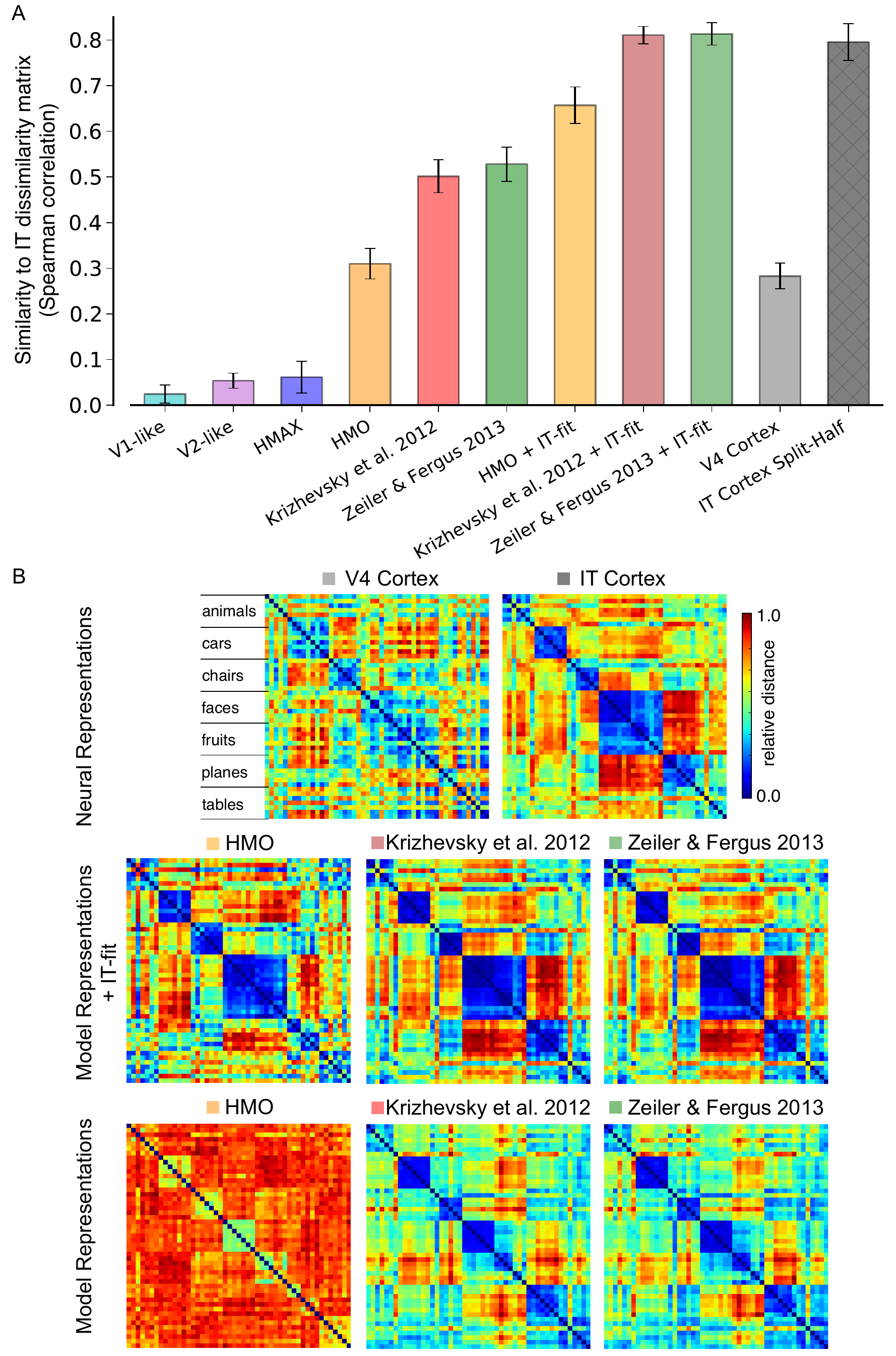}
\end{center}
\caption{{\bf Object-level representational similarity analysis comparing model and neural representations to the IT multi-unit representation.}  A) Following the proposed analysis in~\cite{NikolausKriegeskorte:2008bz}, the object-level dissimilarity matrix for the IT multi-unit representation is compared to the matrices computed from the model representations and from the V4 multi-unit representation.  Each bar indicates the similarity between the corresponding representation and the IT multi-unit representation as measured by the Spearman correlation between dissimilarity matrices.  Error bars indicate standard deviation over 10 splits.  The IT Cortex Split-Half bar indicates the deviation measured by comparing half of the multi-unit sites to the other half, measured over 50 repetitions.  The V1-like, V2-like, and HMAX representations are highly dissimilar to IT cortex.  The HMO representation produces comparable deviations from IT as the V4 multi-unit representation while the Krizhevsky et al. 2012 and Zeiler \& Fergus 2013 representations fall in-between the V4 representation and the IT cortex split-half measurement.  The representations with an appended ``+ IT-fit'' follow the methodology in~\cite{Yamins:2014gi}, which first predicts IT multi-unit responses from the model representation and then uses these predictions to form a new representation (see text).  B) Depictions of the object-level RDMs for select representations.  Each matrix is ordered by object category (animals, cars, chairs, etc.) and scaled independently (see color bar).  For the ``+ IT-fit'' representations, the feature for each image was averaged across testing set predictions before computing the RDM (see Methods).}
\label{fig:nklike}
\end{figure}
Finally, we measured representational similarity using the analysis methodology proposed in~\cite{NikolausKriegeskorte:2008bz}.  This analysis methodology measures how similar two representations are and is robust to global scalings and rotations of the representational spaces.  To compute the representational similarity between the IT multi-unit and model representations, we computed object-level representational dissimilarity matrices (RDMs) for model and neural representations (matrices are 49x49 dimensional as there are 49 total objects).  We then measured the Spearman rank correlations between the model derived RDM and the IT multi-unit RDM (see Methods).  In Figure~\ref{fig:nklike}A we show the results of the representational similarity measurements for the model representations and in Figure~\ref{fig:nklike}B we show depictions of the RDMs for select representations.  For comparison we present the result between the V4 multi-unit representation and the IT multi-unit representation.  To determine the variability due to the IT neural sample, we also present the similarity measurement between one-half of the IT multi-units and the other half (IT Cortex Split-Half).  In addition, we provide results following the methodology in~\cite{Yamins:2014gi}, which first predicts the IT multi-unit site responses from the model representation and then uses these predictions to form a new representation.  We refer to these representations with an appended ``+ IT-fit''.  Our measurements of the HMO + IT-fit representation are in general agreement with the results in~\cite{Yamins:2014gi} but vary slightly because of differences in the image set used to produce these measurements and details of the methodology used to produce the IT predictions.  Interestingly, by fitting a linear transform at the image-level to IT multi-units, the Krizhevsky et al. 2012 and Zeiler \& Fergus 2013 DNNs fall within the noise limit of the IT split-half object-level RDM measurement.  However, the HMO, Krizhevsky et al. 2012, and Zeiler \& Fergus 2013 representations, without the added linear mapping, have deviations from the IT representation that are unexplained by noise variation.  While it is informative that a linear mapping can produce RDMs in correspondence with the IT RDM, we conclude that there remains a gap between DNN models and IT representation when measured with object-level representational similarity.

\section*{Discussion}
In summary, our measurements indicate that the latest DNNs rival the representational performance of IT cortex on a rapid object category recognition task.  We evaluated representational performance using a novel kernel analysis methodology, which measures precision as a function of classifier complexity.  Kernel analysis allows us to measure a desirable property of a representation: a \emph{good} representation is highly performant with a simple classification function and can thus accurately predict class labels from few examples, while a \emph{poor} representation is only performant with complex classification functions and thus requires a large number of training examples to accurately predict (see Methods for elaboration on this point).  Importantly, we made comparisons between models and neural measurements by correcting the models for experimental limitations due to sampling, noise, and trials.  In this analysis we found that the Zeiler \& Fergus 2013 DNN achieved comparable representational performance to the IT cortex multi-unit representation and both the Krizhevsky et al. 2012 and Zeiler \& Fergus 2013 representations surpassed the performance of the IT cortex single-unit representation.  These results reflect substantial progress of computational object recognition systems since our previous evaluations of model representations using a similar object recognition task \cite{Pinto:2008gj,Pinto:2011iw}.  These results extend our understanding over recent, complimentary studies, which have examined representational similarity~\cite{Yamins:2014gi}, by evaluating directly absolute representational performance for this task.  In contrast to the representational performance results, all models that we have tested failed to capture the full explainable variation in IT responses (Figures~\ref{fig:itfit} and~\ref{fig:nklike}).  Nonetheless, our results, in conjunction with the results in Yamins et al. 2014~\cite{Yamins:2014gi}, indicate that the latest DNNs provide compelling models of primate object recognition representations that predict neural responses in IT cortex~\cite{Yamins:2014gi} and rival the representational performance of IT cortex.

To address the behavioral context of core visual object recognition our neural recordings were made using 100 ms presentation times.  We chose only a single presentation time (as opposed to rerunning the experiment at different presentation times) to maximize the number of images and repetitions per image given time and cost constraints in neurophysiological recordings.  This choice is justified by previous results that indicate human subjects are performant on similar tasks with just 13 ms presentation times~\cite{Potter:2013go}, that human performance on similar tasks rapidly increases from 14 ms to 56 ms and has diminishing returns between 56 ms and 111 ms~\cite{Keysers:2001ff}, that decoding from IT at 111 ms presentation times achieves nearly the same performance at 222 ms presentation times~\cite{Keysers:2001ff}, that for 100 ms presentation times the first spikes after stimulus onset in IT are informative and peak decoding performance is at 125 ms~\cite{hung2005fast}, and that maximal information rates in high-level visual cortex are achieved at a rate of 56 ms/stimulus~\cite{Endres:2007iy}.  Furthermore, we have measured human performance on our task and observed that the mean response accuracy at 100 ms presentation times is within 92\% of the accuracy at 2000 ms presentation times (see Figure~\ref{fig:human_perf}).  While reducing presentation time below 50 ms likely would lead to reduced representational performance measurements in IT (see \cite{Keysers:2001ff}), the presentation time of 100 ms we used for our evaluation is applicable for the core recognition behavior, has previously been shown to be performant behaviorally and physiologically, and in our own measurements on this task captures the large majority of long-presentation-time (2 second) human performance.

The images we have used to define the computational task allow us to precisely control variations to object exemplar, geometric transformations, and background; however, they have a number of disadvantages that can be improved upon in further studies.  For example, this image set does not expose contextual effects that are present in the real world and may be used by both neural and machine systems, and it does not include other relevant variations, e.g. lighting, texture, natural deformations, or occlusion.  We view these current disadvantages as opportunities for future datasets and neural measurements, as the approach taken here can naturally be expanded to encompass these issues.

There are a number of issues related to our measurement of macaque visual cortex, including viewing time, behavioral paradigm, and mapping the neural recording to a neural feature, that will be necessary to address in determining the ultimate representational measurement of macaque visual cortex.  The presentation time of the images shown to the animals was intentionally brief (100 ms), but is close to typical single-fixation durations during natural viewing ($\sim$200 ms), and human behavioral testing (Figure \ref{fig:human_perf}) shows that the visual system achieves high performance at this viewing time.  It will be interesting to measure how the neural representational space changes with increased viewing time and multiple fixations.  Another aspect to consider is that during the experimental procedure, animals were engaged in passive viewing and human subjects were necessarily performing an active task.  Does actively performing a task influence the neural representation?  While several studies report that such effects are present, but weak at the single-unit level \cite{Vogels:1995cs,Koida:2006jp,Suzuki:2006ig,OpdeBeeck:2010fe}, no study has yet examined the quantitative impact of these effects at the population level for the type of object recognition task we examined.  Active task performance may be related to what are commonly referred to as attentional phenomena [e.g. biased competition].  In addition, the mapping from multi-unit and single-unit recordings to the neural feature vector we have used for our analysis is only one possible mapping, but it is a parsimonious first choice.  Finally, visual experience or learning may impact the representations observed in IT cortex.  Interestingly, the macaques involved in these studies have had little or no real-world experience with a number of the object categories used in our evaluation, though they do benefit from millions of years of evolution and years of postnatal experience.  However, significant learning effects in adult IT cortex have been observed~\cite{Kobatake:1998vc,Baker:2002bt,Sigala:2002ta}, even during passive viewing~\cite{Li:2010it}.  We have examined the performance of computational algorithms in terms of their absolute representational performance.  It is also interesting to examine the necessary processing time and energy efficiency of these algorithms in comparison to the primate visual system.  While a more in depth analysis of this issue is warranted, from a ``back-of-the-envelope'' calculation (see SI) we conclude that model processing times are currently competitive with primate behavioral reaction times but model energy requirements are 2 to 3 orders of magnitude higher than the primate visual system.

How do our measurements of representational performance relate to overall system performance for this task?  Measuring representational performance fundamentally relies on a measure of the representation, which we have assumed is a neural measure such as single-unit response or multi-unit response.  This poses difficulties for obtaining an accurate measure of human representational performance.  Using only behavioral measurements the representation must be inferred, which may be possible through an investigation of the psychological space of visually presented objects.  However, more direct methods may be fruitful using fMRI (see \cite{Kriegeskorte:2008vz}), or a process that first equates macaque and human performance and uses the macaque neural representation as a proxy for the human neural representation.  One approach to directly measure the overall system performance is to replicate the cross-validated procedure used to measure models in humans.  Such a procedure should control the human exposure to the training set and provide the correct labels on the training set.  The procedure for measuring human performance presented in the SI does not follow this procedure.  However, a comparison between human performance at 100 ms presentation times (see Figure~\ref{fig:human_perf}) and overall DNN model performance on the test-set (see Figure~\ref{fig:svm_absolute}) indicates that there is likely a gap between human performance (85\% mean accuracy) and DNN performance (77\% mean accuracy) on this task because allowing the human subjects exposure to 80\% of the images with the correct labels is only likely to increase the human performance number.  Furthermore, there is individual variability in the human performance with some individuals performing well above the mean.  Therefore, while we have not attempted to make a direct comparison between human performance and DNN performance, we infer that human performance exceeds current DNN performance.

Our methodology and approach relates to the encoding and decoding approaches in systems neuroscience, which, in our view, provide complementary insights into neural visual processing.  The kernel analysis methodology we use here is a neural decoding approach because it measures the relationship between the neural (or model) representation and unobserved characteristics of the stimuli (class labels).  The linear-SVM methodology is also a decoding approach because it tests the generalization performance of predicting the unobserved class label from the neural (or model) representation.  The approaches of predicting IT multi-unit response (Figure~\ref{fig:itfit}) and measuring representational similarity to IT representation (Figure~\ref{fig:nklike}) are encoding approaches because they measure the relationship between functions or measurements derived from the stimuli (pixels in the images) and the neural variation present in IT.  The complementary nature of these approaches is demonstrated in our results.  For example, while the Zeiler \& Fergus 2013 DNN rivals the decoding performance of IT cortex, it fails to capture over 40\% of the explainable variance in the IT neural sample and therefore does not produce a complete neural encoding model.  Conversely, the V4 multi-unit representation severely underperforms the DNNs and IT cortex when measured with decoding approaches but produces comparable results to these representations when predicting IT multi-unit responses with an encoding approach.  It is currently unclear what variation in the IT cortex multi-unit representation is not captured by DNNs or the V4 multi-unit representation.  Furthermore, the IT variation that is captured by DNNs and V4 is, relative to correlations between DNN models, weakly correlated (Figure~\ref{fig:itfit}B).  Overall, the remaining unexplained IT variation may be exposed through a decoding approach (by, for example, exploring additional task labels), through an encoding approach (by exploring additional stimulus transformations), or through approaches that take into account intrinsic neural dynamics (e.g. \cite{Stevenson:2012ux}).  The comparably high performance of the V4 multi-unit representation at predicting IT multi-unit responses may be due to its ability to capture intrinsic neural dynamics present in IT that are unrelated to stimulus derived variables.

It may be surprising that multi-units outperform single-units on this task (see Figure~\ref{fig:multi_vs_single} for trial corrected performance); however, when considering generalization performance, it is not the case that averaged single-units should under perform the original single-units.  Multi-units may provide a form of regularization that is appropriate for this task.  This regularization may be due to averaging out single-unit noise, and/or reducing variation in the representational space that is irrelevant, and therefore spurious, for the task.  Alternatively, the single-unit variation may be appropriate for different tasks that we have not measured, such as fine distinctions between objects or 3-dimensional object surface representation.  Could the regularizing properties of multi-unit responses be indicative of broader regularization mechanisms related to spatial organization in cortex (topographic maps, and functional clustering)?  Just as our multi-unit recordings average together a number of single-units, neurons ``reading-out'' signals from IT cortex could average over cortical topography and thus regularize the classification decision boundary.  This leads to a broader computational goal of finding an appropriate mapping of biological phenomenology and physical mechanism to the computational concepts of kernels, regularizers, and representational spaces.  The overall performance of learning algorithms strongly depends on the interconnections between the choice of kernel, regularizer, and representation.  In our current work (and the predominant mode in the field), we have made specific choices on the kernel and regularizer and examined aspects of the representational space.  However, a full account of biological learning and decision making must determine accurate descriptions for all three of these computational components in biological systems.

Interestingly, many of the computational concepts utilized in the high performing DNNs that we have measured extend back to early models of the primate visual system.  All three DNNs we examined \cite{Krizhevsky:2012wl,Zeiler:2013ws,Yamins:2014gi} implement architectures containing successive layers of operations that resemble the simple and complex cell hierarchy first described by Hubel and Wiesel~\cite{DHHubel:1962vm,hubel1968receptive}.  In particular, max-pooling was proposed in~\cite{Riesenhuber1999} and is a prominent feature of all three DNNs.  Additional computational concepts are convolution or weight sharing, which was introduced in \cite{LeCun:1998hy} and utilized in \cite{Krizhevsky:2012wl,Zeiler:2013ws}, and backpropagation \cite{Rumelhart:1986we}, which is utilized in \cite{Krizhevsky:2012wl,Zeiler:2013ws}.

The success of the Krizhevsky et al. 2012 and the Zeiler \& Fergus 2013 DNNs raises a number of interesting questions.  The categories we used for testing (the 7 classes used in the kernel analysis measurements) are a small fraction of the 1000 classes that these models were trained on, and it is not clear if there is a direct correspondence between the classes in the two image sets.  At this point it is not clear how the non-relevant classes in the set used to train the models affects our performance estimate.  As more detailed analyses are conducted it will be interesting to determine which categories are necessary to replicate ventral stream performance and similarity.  For example, there may be biases in the necessary category distribution toward ecologically relevant categories, such as faces.  Of biological relevance, it is not clear if natural primate development is comparable to the 15M labeled images used to train these DNNs and it seems likely that more innate knowledge of the visual world (acquired during evolution) and/or more unsupervised training (during development) are utilized in biological systems.  Finally, given their similar architectures, it is unclear why some DNNs perform well and others do not.  However, our analyses provide cursory evidence that models with more layers perform better and models that effectively reduce the dimensionality of the original problem perform better.  More work is necessary to determine best practices using these architectures and to determine the importance of hierarchical representations and representations that reduce dimensionality.  The principled approach we have provided here allows for practical evaluations between models and neurons, and may provide a tool in assessing progress in the development of DNNs.  Going forward, we would ideally like a better theoretical understanding of these architectures that would lead to more consistent implementations and would produce a detailed, mechanistic hypothesis for ventral stream processing (see \cite{Mallat:2012by} for an example of such an effort).
\section*{Methods}
\subsection*{Ethics statement}
This study was performed in strict accordance with the recommendations in the Guide for the Care and Use of Laboratory Animals of the National Institutes of Health.  All surgical and experimental procedures were approved by the Massachusetts Institute of Technology Committee on Animal Care (Animal protocol: 0111-003-014).  All human behavioral measurements were approved by the Massachusetts Institute of Technology's Committee on the Use of Humans as Experimental Subjects (number: 0812003043).

\subsection*{Image dataset generation} \label{sec:imageset}
%
%
Synthetic images of objects were generated using POV-Ray, a free, high-quality ray tracing software package (\url{http://www.povray.org}).  3-d models (purchased from Dosch Design and TurboSquid) were converted to the POV-Ray format. This general approach allowed us to generate image sets with arbitrary numbers of different objects, undergoing controlled ranges of identity preserving object variation/transformation. The 2-d projection of the 3-d model was then combined with a randomly chosen background. In our image set, no two images had the same background, in some cases the background was, by chance, correlated with the object (plane on a sky background, car on a street) but more often they were uncorrelated, giving no information about the identity of the object.  A circular aperture with radial fall-off was applied to create each final image.

\subsection*{Neural data collection}
We collected neural data from V4 and IT across two adult male rhesus monkeys (\textsl{\small{Macaca mulatta}}, 7 and 9 kg) by using a multi-electrode array recording system (BlackRock Microsystems, Cerebus System).  We chronically implanted three arrays per animal and recorded the 128 most visually driven neural measurement sites (determined by separate pilot images) in one animal (58 IT, 70 V4) and 168 in another (110 IT, 58 V4).  During image presentation we recorded multi-unit neural responses to our images from the V4 and IT sites.  Images were presented on an LCD screen (Samsung, SyncMaster 2233RZ at 120Hz) one at a time.  Each image was presented for 100 ms with a diameter of 8$^{\scriptstyle\circ}$ (visual angle) at the center of the screen on top of the half-gray background and was followed by a 100 ms half-gray ``blank'' period.  The animal's eye position was monitored by a video eye tracking system (SR Research, EyeLink II), and the animal was rewarded upon the successful completion of 6--8 image presentations while maintaining good eye fixation (within $\scriptstyle \pm$2$^{\scriptstyle\circ}$) at the center of the screen, indicated by a small (0.25$^{\scriptstyle\circ}$) red dot.  Presentations with larger eye movements were discarded.  In each experimental block, we recorded responses to all images.  Within one block each image was repeated once.  Over all recording sessions, this resulted in the collection of 47 image repetitions, collected over multiple days.  All surgical and experimental procedures were done in accordance with the National Institute of Health guidelines and the Massachusetts Institute of Technology Committee on Animal Care.

To arrive at the multi-unit neural representation, we converted the raw multi-unit neural responses to a neural representation through the following normalization process.  For each image in a block, we compute the vector of raw firing rates across measurement sites by counting the number of spikes between 70 ms and 170 ms after the onset of the image for each site.  We then subtracted the background firing rate, which is the average firing rate during presentation of a gray background (``blank'' image), from the evoked response.  In order to minimize the effect of variable external noise, we normalize each site by the standard deviation of each site's response to a block of images.  Finally, the neural representation is calculated by taking the mean across the repetitions for each image and for each site, producing a scalar valued matrix of neural sites by images.  This post-processing procedure is only our current best-guess at a neural code, which has been shown to quantitatively account for human performance~\cite{Majaj:2012ui}.  Therefore, it may be possible to develop a more effective neural decoding for example influenced by intrinsic cortical variability~\cite{Stevenson:2012ux}, or dynamics~\cite{Churchland:2012bq,Canolty:2010ts}.

To arrive at the single-unit neural representation, we followed a similar normalization process as the multi-unit representation, but first conducted spike-sorting on the multi-unit recordings.  We sorted single-units from the multi-unit IT and V4 data by using affinity propagation~\cite{Frey:2007hs} together with the method described in~\cite{Quiroga:2004hl}.  Using a conservative criteria we isolated 160 single-units from the IT recordings and 95 single-units from the V4 recordings with 6 repetitions per image for each single-unit.  Given these single-unit responses for each image we followed a processing procedure identical to the multi-unit procedure, which included counting the number of spikes between 70 ms and 170 ms after the onset of the image, subtracting the background firing rate, and normalizing by the standard deviation of the site's response to a block of images.  Finally, we selected from these single-units the top 40 based on response consistency over trials on a separate image set.  Specifically, we measured the response to 280 images not included in our evaluations of performance but drawn from a similar stimulus distribution.  These images contained 7 unique objects not contained in the original set with 1 object from each of the 7 categories.  For each object there are 40 images, each with a unique background and with the object position, scale, and pose randomly sampled.  For each single-unit, we separated the responses over trials into two groups, averaged across trials, and measured the correlation of these response vectors.  We then sorted the single-units based on this correlation and selected the 40 with highest correlation and therefore most consistent single-units.  We repeated this procedure separately for V4 and IT.  The resulting consistency measurement for the IT single-units was comparable to other measurements using single-unit electrophysiology~\cite{Rust:2010uk}.  The consistency of the 40 IT single-units was higher than the 15th percentile of single-units measured in~\cite{Rust:2010uk}.  In other words, the least consistent IT single-unit in the group of 40 was more consistent than the bottom 15\% of single-units analyzed in~\cite{Rust:2010uk}.  Unless otherwise noted, all analyses of single-units use these 40 selected single-units.

\subsection*{Kernel analysis methodology}
We would also like to measure accuracy as we change the complexity of the prediction function.  To accomplish this, we use an extension of the work presented in~\cite{Montavon:2011wp}, which is based on theory presented in~\cite{Braun:2006va}, and~\cite{Braun:2008ul}, and we refer to as \emph{kernel analysis}.  We provide a brief description of this measure and refer the reader to those references for additional details and justification on measuring precision against complexity.  This procedure is a derivative of regularized least squares~\cite{Rifkin:2007wt}, which arrises as a Tikhonov minimization problem~\cite{Evgeniou:2000fx}, and can be viewed as a form of Gaussian process regression~\cite{Rasmussen:2006vz}.  We do not elaborate on the relationships between these views, but each is a valid interpretation on the procedure.

The measurement procedure, which we refer to here as \emph{kernel analysis}, utilizes regularized kernel ridge regression to determine how well a task in question can be solved by a regularized kernel.  A highly regularized kernel will allow only a simple prediction function in the feature space, and a weakly regularized kernel will allow a complex prediction function in the feature space.  A \emph{good} representation will have high variability in relation to the task in question and can effectively perform the task even under high regularization.  Therefore, if the eigenvectors of the kernel with the largest eigenvectors are effective at predicting the categories, a highly regularized kernel will still be effective for that task.  In contrast, an ineffective representation will have very little variation relevant for the task in question and variation relevant for the task is only contained in the eigenvectors corresponding to the smallest eigenvalues of the kernel and only a weakly regularized kernel will be capable of performing the task efficiently.  Changing the amount of regularization changes the complexity of the resulting decision function: highly regularized kernels allow for only simple decision functions in the feature space and weakly regularized kernels allow for complex decision functions.  Intuitively, a good representation is one that learns a simple boundary (highly regularized) from a small number of randomly-chosen examples, while a poor representation makes a more complicated boundary (weakly regularized), requiring many examples to do so.

In our formulation, kernel analysis consists of solving a regularized least squares or equivalently a kernel regression problem over a range of regularization~\cite{Rifkin:2007wt}.  The regularization parameter $\lambda$ controls the complexity of the function and the precision or performance is measured as the leave-one-out generalization error ($looe$).  We refer to the inverse of the regularization parameter ($1/\lambda$) as the \emph{complexity} and $1 - looe(\lambda)$ as the \emph{precision}.  Thus, the curve $1 - looe(\lambda)$ provides us with a measurement of the precision as a function of the model complexity for the given representational space.  The curves produced by different representational spaces will inform us about the simplicity of the task in that representational space, with higher curves indicating that the problem is simpler for the representation.

One of the advantages of kernel analysis is that the kernel PCA method converges favorably from a limited number of samples.  \cite{Braun:2008ul} shows that the kernel PCA projections obtained with a finite and typically small number of samples $n$ (images in our context) are close with multiplicative errors to those that would be obtained in the asymptotic case where $n \mapsto \infty$.  This result is especially important in our setting as the number of images we can reasonably obtain from the neural measurements is comparatively low.  Therefore, kernel analysis provides us with a methodology for assessing representational effectiveness that has favorable properties in the low image sample regime, here thousands of images.

We next present the specific computational procedure for computing kernel analysis.  Given the learning problem $p(x,y)$ and a set of $n$ data points $\{(x_1,y_1),..., (x_n,y_n)\}$ drawn independently from $p(x,y)$ we evaluate a representation defined as a mapping $x\mapsto \phi(x)$.  For our case, the inputs $x$ are images, the $y$ are normalized category labels, and the $\phi$ denotes a feature extraction process.  

As suggested by~\cite{Montavon:2011wp}, we utilize the Gaussian kernel because this kernel implies a smoothness of the task of interest in the input space~\cite{Smola:1998dq} and does not bias against representations that may be more adapted to non-linear regression functions.  We compute the kernel matrix $K_\sigma$ associated to the data set as
\begin{equation}
K_\sigma =  \begin{pmatrix}
k_\sigma(\phi(x_1),\phi(x_1)) & ... & k_\sigma(\phi(x_1),\phi(x_n)) \\
 \vdots &  &\vdots \\
k_\sigma(\phi(x_n),\phi(x_1)) & ... & k_\sigma(\phi(x_n),\phi(x_n)) \\
\end{pmatrix},
\end{equation}
where the standard Gaussian kernel is defined as $k_\sigma (x, x') = \exp ( - || x - x' ||^2 / 2\sigma^2)$ with scale parameter $\sigma$.

The regularized kernel regression problem is
\begin{equation}
\operatorname{min}_{\Theta \in R^n} \frac{1}{2} \sum^n_{i=1} \| Y - K_\sigma \Theta \|^2_2 + \frac{\lambda}{2} \Theta^t K_\sigma \Theta,
\label{eq:loss}
\end{equation}
where $Y$ are the normalized vector of category labels, $\Theta$ is the vector of regression parameters and $\lambda$ is the regularization scalar.  The solution to the regularized regression problem for a fixed $\sigma$ and $\lambda$ is denoted as $\Theta_{\sigma}(\lambda)$ and is given as
\begin{equation}
\Theta^*_{\sigma}(\lambda) = (K_\sigma + \lambda I)^{-1} Y,
\label{eq:estimator}
\end{equation}
where $I$ is the identity matrix.

The leave-one-out error can be calculated as
\begin{equation}
LOOE_{\sigma}(\lambda) = \frac{\Theta^*_{\sigma}(\lambda)} {diag((K_\sigma + \lambda I)^{-1})},
\end{equation}
where $diag(M)$ denotes the column vector satisfying $diag(M)_i = M_{ii}$ and the division is elementwise.  Note that $LOOE_{\sigma} (\lambda)$ is a vector of errors for each example and we compute the mean squared error over the examples as $looe_{\sigma}(\lambda) = \frac{1}{n}\sum_i (LOOE_{\sigma}(\lambda)_{i})^2$, which is considered a good empirical proxy for the error on future examples (generalization error).  We note that the leave-one-out error can be computed efficiently given an eigendecomposition of the kernel matrix.  We refer to \cite{Rifkin:2007wt} for derivation and details on this computation.

To remove the dependence of the kernel on $\sigma$ we find the value that minimizes $looe_{\sigma}(\lambda)$ at that value of $\lambda$: $looe(\lambda) = \operatorname{min}_{\sigma} looe_{\sigma} (\lambda)$.  Finally, for convenience we plot \emph{precision}\, ($1 - looe(\lambda)$) against \emph{complexity}\, ($1 / \lambda$).  Note that because we optimize the value of $\sigma$ for each value of complexity, $\sigma$ (the width of the Gaussian kernel) will regularize the feature space when $\lambda$ does not provide sufficient regularization.  Because of this effect, the precision-complexity curves in Figure \ref{fig:ka_models} plateau at high values of complexity because the optimal value of $\sigma$ increases in the high complexity regime (low values of $\lambda$).  We would otherwise expect that at high complexity, the regression would over fit and produce poor generalization (low $1 - looe(\lambda)$ or high error).  At low values of complexity ($\lambda$) we observe that there is an optimal value for $\sigma$, which is dependent on the representation, and is robust over image splits.  Therefore, it is not possible to achieve high precision at low complexity simply by reducing $\sigma$; the variation in the representational space must be aligned with the task in order to achieve high precision at low complexity.

%
Note that we have chosen to use a squared error loss function for our multi-way classification problem.  While it might be more appropriate to evaluate a multi-way logistic loss function, we have chosen to use the least-squares loss because it provides a stronger requirement on the representational space to reduce variance within category and to increase variance between categories, and it allows us to distinguish representations that may be identical in terms of separability for a certain complexity but still have differences in their feature mappings.  The kernel analysis of deep Boltzmann machines in~\cite{Montavon:2012ub} also uses a mean squared loss function in the classification problem setting and is widely used in machine learning.

In the discussion above, $Y = (y_1, \ldots, y_n)$ represents the vector of task labels for the images $(x_1, \ldots, x_n)$.  In our specific case, the $y_i$ are normalized category identity values (normalized such that predicting 0 will result in a precision equal to 0 and perfect prediction will result in a precision equal to 1).  To generalize to the case of multiway categorization, we use a version of the common one-versus-all strategy.  Assuming $k$ distinct categories, for each category $j$ we compute the per-class leave-one-out error $looe_{\sigma} (\lambda)^j$ by replacing $Y$ in equations \ref{eq:loss} and \ref{eq:estimator} with $Y_j$.  The overall leave-one-out error is then the average over categories of the per-category leave-one-out error.  Minimization over $\sigma$ then proceeds as in the single category case.

To evaluate both neural representations and machine representations using kernel analysis we measure the $1 - looe(\lambda)$ curve.  The image dataset consists of 1960 images containing seven object categories with seven instances per object category.  The categories are Animals, Cars, Chairs, Faces, Fruits, Planes and Tables.  To measure statistical variation due to subsampling of image variation parameters we compute the $1 - looe(\lambda)$ curve ten times, each time sampling 80\% of the images with replacement.  The ten samples are fixed for all representations and within each subset we equalize the number of images from each category.  For each representation, we maximize over the values of the Gaussian kernel $\sigma$ parameter as follows.  We evaluate the Gaussian kernel scale parameter $\sigma$ for each representation at a range of values centered on the median distance over the distribution of distances for that representation.  With $\sigma_m$ denoting the value equal to the median distance, we evaluated kernels with $\sigma = \alpha \sigma_m$ for 32 values of $\alpha$ in the range of $0.1$ to $10$ with logarithmic spacing.  We found that in practice the values of $\sigma$ that minimized the leave-one-out error were close to $\sigma_m$ and well within this range.  To span a large range of complexity we evaluate $looe(\lambda)$ at 56 values of $\lambda$ from $10^{-4}$ to $10^3$ with logarithmic spacing.  For each representation, this procedure produces a curve for each of the subsets, where the mean and spread (range of the values over the 10 subsets for each value of $\lambda$) are shown in Figure \ref{fig:ka_models}.  As a summary value, we also compute the area under the curve (KA-AUC) for each of the image set randomizations and report the mean and standard deviation in Table 1.  We use these mean KA-AUC values for the measurements in Figures \ref{fig:subsampled}A and \ref{fig:subsampled}B.

\subsection*{Machine representations}
We evaluate a number of model representations from the literature, including several recent best of breed representational learning algorithms and visual representation models.  In particular we examine three recent convolutional DNNs~\cite{Krizhevsky:2012wl,Zeiler:2013ws} + Yamins.  The DNNs in \cite{Krizhevsky:2012wl,Zeiler:2013ws} are of note because they have each successively surpassed the state-of-the-art performance on the ImageNet Large Scale Visual Recognition Challenge (ILSVRC).  Note that results have continued to improve on this challenge since we ran our analysis.  See \url{http://www.image-net.org/} for the latest results.

Given the similarity of these DNNs to models of biological vision and our particular interest in the primate visual system, we also tested two representations that attempt to capture ventral stream processing.  The ``V1-like'' model attempts to replicate functional responses in the primary visual area, the ``V2-like'' model similarly replicates response properties of the secondary visual area, and the HMAX instantiation is a model of ventral visual processing that implements the simple and complex cell hierarchy proposed in~\cite{Riesenhuber1999}.

\textbf{V1-like~\cite{Pinto:2008gj}} We evaluate the V1-like representation from Pinto et al.'s V1S+~\cite{Pinto:2008gj}.  This model attempts to capture a first-order account of primary visual cortex (V1).  It computes a collection of locally-normalized, thresholded Gabor wavelet functions spanning orientation and frequency.  This model is a simple, baseline biologically-plausible representation, against which more sophisticated representations can be compared.  This model has 86400 dimensions.

\textbf{V2-like~\cite{Freeman:2011gl}} We evaluated a recent proposal for the functional role of visual area V2~\cite{Freeman:2011gl}.  This model constructs a representation from conjunctions of Gabor outputs, which is similar to the V1-like model.  The Gabor outputs are combined non-linearly and averaged within receptive field windows.  The representation formed by this model has been shown to correspond to visual area V2 and explains visual crowding~\cite{Freeman:2011gl}.  The output representation for the V2-like model has 24316 dimensions.

\textbf{HMAX instantiation~\cite{Mutch:2008eo}}  We evaluate the model in \cite{Mutch:2008eo}, which is a biologically inspired hierarchical model utilizing sparse localized features.  This model has been shown to perform relatively well on previous measures of invariant object recognition \cite{Pinto:2009ho}, and to explain some aspects of ventral stream responses \cite{Riesenhuber1999,Serre:2007jy}.  This representation has 4096  dimensions.  Counting the simple-complex module as a single layer, this model has two layers.

\textbf{HMO, Yamins et al. 2014~\cite{Yamins:2014gi}} The model uses hierarchical modular optimization (HMO) to develop a large, deep network that is a combination of convolutional neural networks and is described in~\cite{Yamins:2014gi}.  The HMO algorithm is an adaptive boosting procedure that interleaves hyper-parameter optimization (see \cite{Yamins:2014gi} and references therein).  The specific model we examine here is identical to the one in Yamins et al. 2014~\cite{Yamins:2014gi}.  This model is developed on a screening task that contains images of objects placed on randomly selected backgrounds.  This task is similar in its construction to the task we use here to evaluate representational performance, however, it contains entirely different objects in totally non overlapping semantic categories, with none of the same backgrounds and widely divergent lighting conditions and noise levels.  The specific model we examine was produced by the HMO optimization procedure on this screening task and is a convolutional neural network with 1250 top-level outputs.  Therefore, the total dimensionality of the HMO representation is 1250 and the model is composed of four layers.  The model is identical to the one presented in Yamins et al. 2014~\cite{Yamins:2014gi}.  However, the analysis methodology in \cite{Yamins:2014gi} has a number of important differences to the analyses presented in this paper.  The following differences are noteworthy: in this paper we examine only the high variation task and a subset of the images in the high variation task (7 categories with 7 objects per category in this paper vs. 8 categories with 8 objects per category in \cite{Yamins:2014gi}), and in this paper we compute representational similarity on the raw HMO features, while in~\cite{Yamins:2014gi} representational similarity is calculated on the HMO-based IT model population.

\textbf{Krizhevsky et al. 2012~\cite{Krizhevsky:2012wl} (SuperVision)} We evaluate the deep convolutional neural network model ``SuperVision'' described in~\cite{Krizhevsky:2012wl}, which is trained by supervised learning on the ImageNet 2011 Fall release ($\sim$15M images, 22K categories) with additional training on the LSVRC-2012 dataset (1000 categories).  The authors computed the features of the penultimate layer of their model (4096 features) on the testing images by cropping out the center 224 by 224 pixels (this is the input size to their model).  This mimics the procedure described in~\cite{Krizhevsky:2012wl}, in which this feature is fed into logistic regression to predict category labels.  This model has seven layers as tested (counting a layer for each linear-convolution or fully connected dot-product).

\textbf{Zeiler and Fergus 2013~\cite{Zeiler:2013ws}} The model is a very large convolutional network with 8 layers~\cite{Zeiler:2013ws} trained using supervised learning on the LSVRC-2012 dataset to predict 1000 output categories. The training data is augmented by taking random crops and flips out of the 256 by 256 pixel images as in ~\cite{Krizhevsky:2012wl} to prevent overfitting. Additionally, visualization experiments exploring what deep models learn lead to hyperparameter choices which yield better performance.  The feature representation is the 4096 features input to the softmax classifier averaged over crops from the 4 corners, center, and horizontal flips (producing ten 4096 dimensional vectors that are averaged to produce the 4096 dimensional representation used here).  This model also has seven layers.

\subsection*{Experimental noise matched model}
In our evaluation of representational performance we are limited by the observed noise in the neural representation.  To produce a fair comparison we alter the model representational measurements by adding a level of noise that is matched to that observed in the neural representation.  Note that we are unable to fully remove noise from the neural representation, and therefore we add noise to the model representations.  The sources of the observed neural noise are various and we do not make an attempt to distinguish the sources of noise.  Broadly, these noise sources are experimental in nature (e.g. physical electrode movement over time) or are intrinsic to the system.  Examples of noise variation that are intrinsic to the system include the arousal state of the animal, trial-to-trial variability of neural responses, and correlated neural activity that is intrinsic to the system and not related to the stimulus condition.

We estimate a rate-dependent additive noise model from either the multi-unit or single-unit neural responses.  The neural responses entering this noise model estimation follow the same averaging and background subtraction step as described previously.  The rate-dependent additive noise model follows the common observation that spike counts of neurons are approximately Poisson~\cite{Tolhurst:1983wa,Shadlen:1998ta}.  We first normalize the variance of the neural response to 1,
\begin{equation}
\frac{1}{NMT}\sum_{ijk} (r_{ijk} - \mu)^2 = 1,
\end{equation}
where $r_{ijk}$ indicates the neural response with indices $i$ over $N$ neural sites, $j$ over $M$ images, and $k$ over $T$ trials, and $\mu$ is the mean over sites, images and trials.  For each neural site we estimate a linear fit of the relationship between the mean response over trials ($\mu_{ij}$) and the variance over trials ($\sigma_{ij}$) as,
\begin{equation}
a_i \mu_{ij} + b_i = \sigma_{ij}
\end{equation}
with coefficients $a_i$ and $b_i$.  We then average these coefficients over the sites to produce a single relationship between mean rate and variance.  For the multi-unit sites we estimate $a=0.14$ and $b=0.92$ and for the single-unit sites we estimate $a=0.76$ and $b=0.71$.

The noise model assumes that the empirically observed response variation is due to an underlying and unobserved signal contribution and a noise contribution.  We can estimate the total variance of the empirical response ($\sigma^2_{signal+noise}$) as,
\begin{equation}
\begin{aligned}
\sigma^2_{signal+noise} &=  \frac{1}{NM} \sum_{ij}\left [ \left (\frac{1}{T}\sum_k r_{ijk} \right) - \mu \right]^2,\\
\sigma^2_{signal+noise} &= \sigma^2_{signal} + \sigma^2_{noise}.
\end{aligned}
\end{equation}
Note that we cannot observe $\sigma^2_{signal}$ and we use the subscript ``noise'' to denote the standard error of the mean of the signal estimate and not ordinary trial to trial variability.  Therefore, the estimate of the standard error of the signal is given as,
\begin{equation}
\sigma^2_{noise} = \left (\frac{1}{NM}\sum_{ij} ( a \mu_{ij} + b)^2\right) / T.
\end{equation}
Note that we estimate the standard error of the mean using the noise model, which has the benefit of being jointly estimated from the population of neural responses.  In other words, if different neural sites share a similar noise model our estimate of that noise model is improved by estimating its parameters from all of the sites jointly.  However, we have verified that using the empirical estimate of $\sigma^2_{noise}$ gives nearly identical results.

To add noise to the model representations we first scale the total variance of the representation such that the variance after adding noise will be approximately equal to the variance observed in the neural sample.  To do this we scale the model representation such that:
\begin{equation}
\frac{1}{LM}\sum_{lj} (s_{lj} - \nu)^2 = \sigma^2_{signal+noise} - \sigma^2_{noise},
\end{equation}
where $s_{lj}$ indicates the model representation value with index $l$ over $L$ feature dimensions and $\nu$ is the mean over image and features for the model representation.  Finally we add signal dependent noise to the model representation:
\begin{equation}
\hat s_{lj} = s_{lj} + \mathcal{N} (0, (a s_{lj} + b)^2 / T),
\label{eq:noisemodel}
\end{equation}
where $\mathcal{N}(\mu,\sigma^2)$ indicates a sample from the normal distribution with mean $\mu$ and variance $\sigma^2$ and $\hat s$ indicates the noise corrupted model representation.  We verified empirically that the resulting variance of the noise corrupted model representation was close to the empirical neural signal variance.  We repeat this procedure for the multi-units and single-units to arrive at model representations that are corrected for the observers noise in the respective neural measurement.

See Figure~\ref{fig:noisetest} for a validation of the experimental noise matched model.

\subsection*{Linear-SVM methodology}
See Supporting Information.

\subsection*{Predicting IT multi-unit sites from model representations}
We performed generalized linear model analysis using ridge regression to predict the IT multi-unit responses from the model representations~\cite{wu2006complete}.  We utilized the same cross-validation procedure as the linear-SVM analysis, estimating encoding models on 80\% of the data and evaluating the performance on the remaining 20\% and repeating this for 10 randomizations.  For each IT multi-unit we estimated an encoding model, predicted the testing responses, and determined the explained explainable variance for that multi-unit.  To measure explainable variance we measured for each multi-unit site and each training set, the Spearman-Brown corrected split-half self-consistency over image presentations (trials).  To arrive at a summary statistics we took the median explained explainable variance over the multi-units for each model representation.  We also used the V4 multi-unit responses to produce encoding models of the IT multi-unit response.  The results of this analysis are presented in Figure~\ref{fig:itfit}.  Note that the predictions produced by the V2-like representation proved to be unstable and produced high variance compared to the other representations.

\subsection*{Representational similarity analysis}
To measure representational similarity we followed the analysis methodology in~\cite{NikolausKriegeskorte:2008bz}.  We first computed a feature vector per object for each representation by averaging the representational vectors over image variations.  The representational dissimilarity matrix (RDM) is defined as
\begin{equation}
RDM(\phi(x))_{ij} = 1 - \frac{cov(\phi(x_i), \phi(x_j))}{\sqrt{var(\phi(x_i))\cdot var(\phi(x_j))}},
\end{equation}
where $\phi(x)$ indicates the representation averaged for each object, $i$ and $j$ index the objects, $cov$ indicates the covariance between the vectors and $var$ the variance of the vector.  Because we have 49 unique objects in our task the resulting RDM is a 49x49 matrix.  To measure the relationship between two RDMs we measured the Spearman rank correlation between the upper-triangular, non-diagonal elements of the RDMs.  We computed the RDM on 20\% of the images and repeated the analysis 10 times.  To compute noise due to the neural sample, we computed the split-half consistency between one half of the IT multi-units and the other half.  We repeated this measurement over 50 random groupings and over the 10 image splits.  Following the methodology in~\cite{Yamins:2014gi}, we also predicted IT multi-unit responses to form a new representation, which we measured using representational similarity analysis.  To produce IT multi-unit predictions for each model representation, we followed the same methodology as described previously (Predicting IT multi-unit sites from model representations).  For each image split, we estimated encoding models on 80\% of the images for each of the 168 IT multi-units and produced predictions for each multi-unit on the remaining 20\% of the images.  We then used these predictions as a new representation and computed the object-level RDM for the 20\% of held-out images.  We repeated the procedure 10 times.  Note that the 20\% of images used for each split was identical for all RDM calculations and that the images used to estimate multi-unit encoding models did not overlap with the images used to calculate the RDM.  The analysis of the representations with the additional IT multi-unit fit can be seen as a different evaluation metric to the results of predicting IT multi-units.  In other words, in Figure~\ref{fig:itfit} we evaluate the IT multi-unit predictions using explained variance at the image-level, and in Figure~\ref{fig:nklike} for the ``+ IT-fit'' representations we evaluate the IT multi-unit predictions using an object-level representational similarity analysis.

\section*{Acknowledgments}
This work was supported by the U.S. National Eye Institute (NIH NEI: 5R01EY014970-09), the National Science Foundation (NSF: 0964269), and the Defense Advanced Research Projects Agency (DARPA: HR0011-10-C-0032).  C.F.C was supported by the U.S. National Eye Institute (NIH: F32 EY022845-01).  We thank Alex Krizhevsky, Matt Zeiler, Quoc Le, and Adam Coates for their help in evaluating their models and comments on the paper.  We also thank Tomaso Poggio, Elias Issa, Christopher K.I. Williams, and Lorenzo Rosasco for their helpful feedback and comments.

\pagebreak
\beginsupplement
\section*{Supporting Information}
\subsection*{Validation of the experimental noise matched model}
\begin{figure*}[t]
\begin{center}
\includegraphics[width=.7\linewidth]{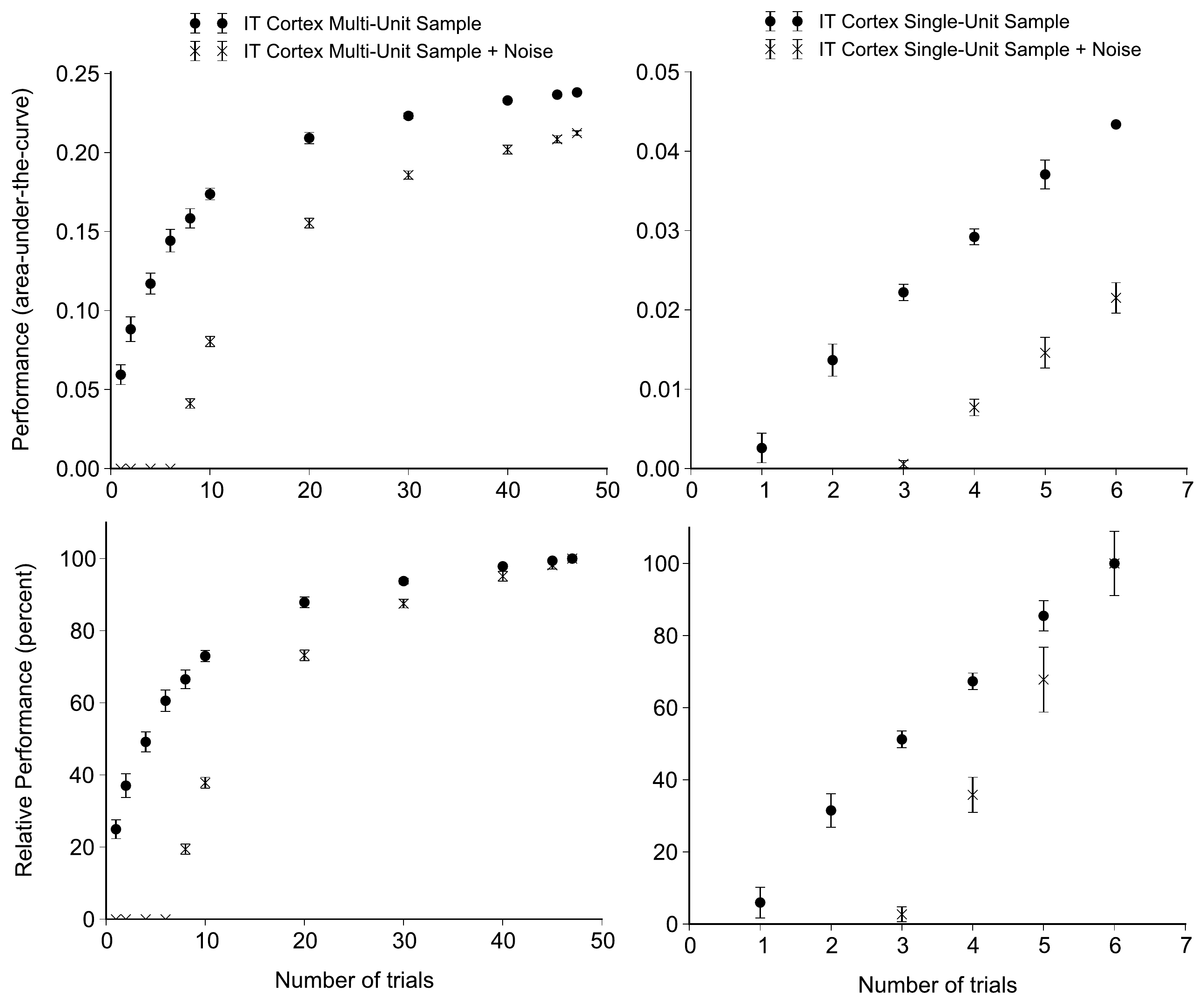}
\end{center}
\caption{{\bf Effects on kernel analysis performance of empirical noise vs. induced noise model.}  In the top left panel we show the performance measurements, as measured by kernel analysis area-under-the-curve, of the IT cortex multi-unit sample and of the IT cortex multi-unit sample with trial dependent added noise as we vary the number of experimental trials (repetitions per image) or the trials in the noise model ($T$ in Eq. \ref{eq:noisemodel}).  In all plots error bars indicate standard deviations of the measure over 10 repetitions of the analysis.  Results are replotted and divided by the maximum performance (Relative Performance) in the lower left panel.  The same analysis is performed for the IT cortex single-unit sample in the right panels.  These results indicate that the noise model reduces our performance measurement over the empirically observed noise and is therefore a conservative model for inducing noise in model representations.  In other words, these results indicate that the model representations with neural matched noise are likely overly penalized.}
\label{fig:noisetest}
\end{figure*}

To determine the effect of experimental and neural variability we measured the result of reducing experimental trials and varying the trials in Eq. \ref{eq:noisemodel} of the noise model.  These results are shown for both multi-unit and single-unit IT cortex samples in Figure~\ref{fig:noisetest}.  The measurements of IT cortex multi-unit sample and single-unit sample show the effect of noise as empirically observed from increasing the number of trials (47 total trials for the multi-unit sample and 6 trials for the single-unit sample).  We measured the effect of our noise model by starting with the IT cortex sample ($s_{lj}$ in Eq. \ref{eq:noisemodel}) at the maximum recorded trials, and added noise according to Eq. \ref{eq:noisemodel} while varying the number of trials in the model ($T$ in Eq. \ref{eq:noisemodel}).  Not surprisingly this produces a reduction in measured performance.  Importantly, the rate of performance decrease is larger in magnitude for the noise model than the observed empirical effect of reducing the number of trials.  This can be seen by examining the relative performance in Figure~\ref{fig:noisetest}.  Therefore, the noise model is a conservative model for inducing noise in model representations because it overly penalizes a representation as we reduce the number of trials or, equivalently, increase the amount of noise.

\subsection*{Human performance on our task as a function of presentation time}
In this section we present measurements of human performance on our task that contribute to our justification for measuring the neural representation at 100 ms presentation times.  A number of criteria influence our choice of presentation time for the neural measurements, which for the following reasons we have chosen to be 100 milliseconds (ms).  First, we seek to address a sub-problem in general visual behavior: core visual object recognition~\cite{DiCarlo:2012em}, or visual information processing within one saccade without contextual influence.  The typical time between saccades is approximately 200-250 ms under natural viewing conditions~\cite{Andrews19992947} and we therefore require a presentation time equal to or less than 200 ms.  Second, to improve the throughput of our neural recordings we desire to minimize the presentation time.  However, in opposition of minimizing the presentation time for throughput, our final criteria is to choose a presentation time for which the primate visual system is still performant.  If we were to reduce the presentation time toward 0 ms we expect the human behavioral performance on this task to approach chance.  Therefore we do not want to reduce the presentation time so much that the primate system fails to perform the relevant visual behavior.  To address this issue, we conducted a psychophysical experiment to evaluate the effect of presentation time on performance for our object recognition task, which we describe next.  In summary, given these three criteria we seek a presentation time that is within the time of a typical saccade, and is long enough that the primate visual system is still performant at that presentation time.
\begin{figure*}[t]
\begin{center}
\includegraphics[width=.6\linewidth]{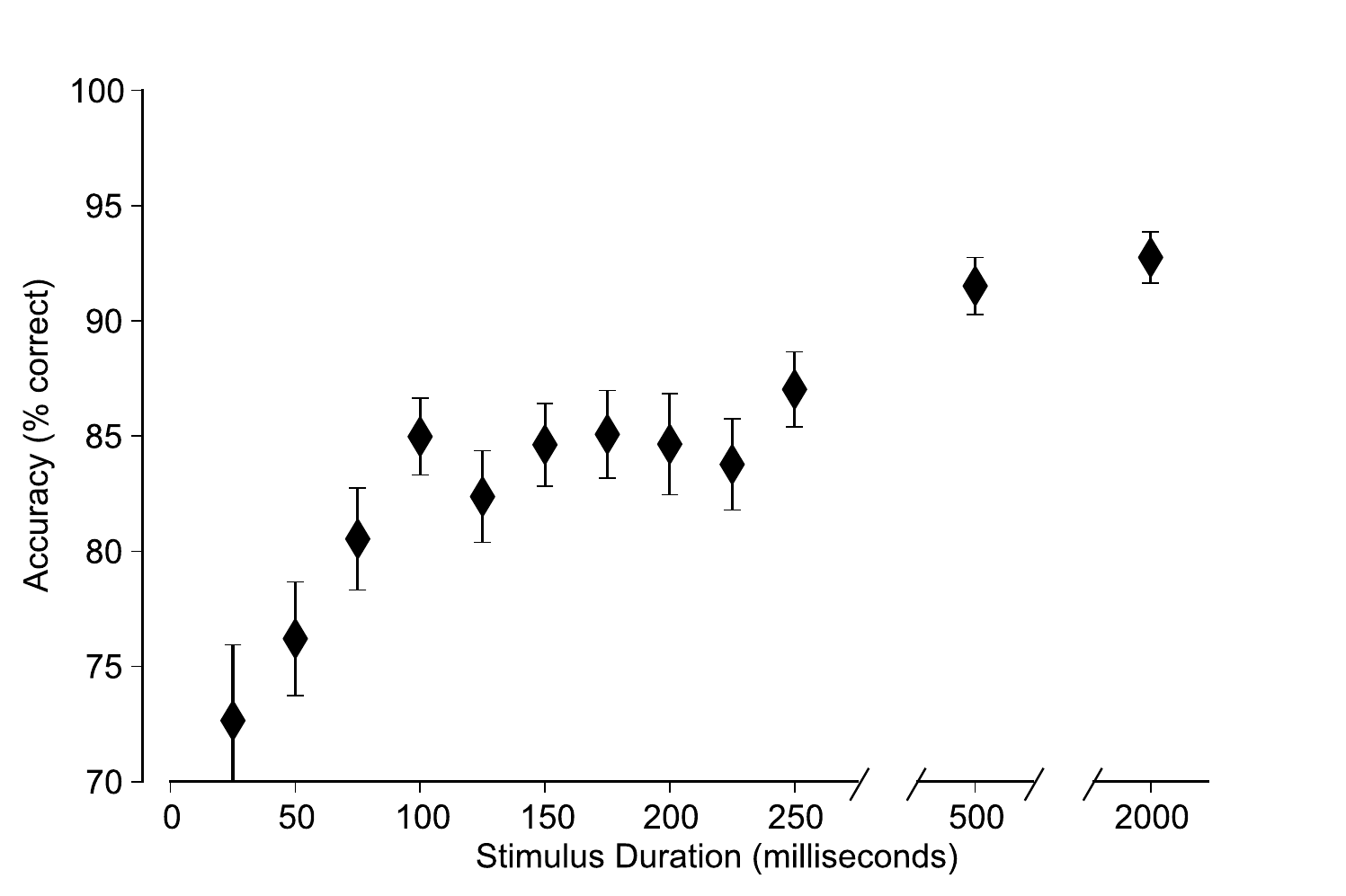}
\end{center}
\caption{{\bf Human performance on the visual recognition task as a function of presentation time.}  We plot the mean block-accuracy for different stimulates presentation durations from responses measured using Amazon Mechanical Turk.  The mean accuracy is plotted as diamond markers and the error bars indicate the 95\% confidence interval of the standard error of the mean over block-accuracies.  Chance performance is $\sim$14\% for this task.  The accuracy quickly increases such that at 100 ms stimulus duration it is within 92\% of the performance at 2 seconds.  This indicates that on this task, human subjects are able to perform relatively highly even during brief presentations of 100 ms.  We refer to this ability as ``core visual object recognition''~\cite{DiCarlo:2012em} and we seek to measure the neural representational performance that subserves this ability.}
\label{fig:human_perf}
\end{figure*}

To justify our selection of the presentation time of 100 ms, we measured human subjects using Amazon Mechanical Turk on our visual category object recognition task for different presentation times (Figure \ref{fig:human_perf}).  The results of this experiment indicate that human behavioral performance reaches a mean accuracy of 92.8\% for 2 second presentation times.  While it might be a priori expected that humans would reach 100\% accuracy for 2 second presentation times, the difficultly of this task results in reduced performance.  From manual examination of typical errors, image instances that have non-prototypical poses of objects (e.g. the underside of a car) or that have the object instance occluded by the image boundary lead to errors.  Regardless of this ceiling effect, we find that human performance is quite robust to reduced presentation times.  We observe that mean performance at 50 ms presentations reaches 82\% of the performance at 2000 ms presentations.  We expect that if we continue to decrease the presentation time the human performance will necessarily approach chance performance of $\sim$14\%.  However, instead of a linear decrease in performance as we reduce the presentation time, we see a saturation effect in performance with the majority of the performance obtained for presentations of 50 ms.  For our chosen presentation time of 100 ms for the neural recordings we observe that mean human performance is 92\% of the 2000 ms presentation time performance.  Furthermore, the performance at 100 ms is close to the performance at 200 ms (the mean time of a saccade).  Therefore, justifying our neural recording presentation time, 100 ms is within typical fixational eye-movements, 100 ms allows us to nearly double our throughput for data collection over 200 ms, and the primate visual system is able to achieve high performance on this task at 100 ms, making it a relevant behavioral regime for study.

The details of the human behavioral measurements are as follows.  We recruited subjects through Amazon Mechanical Turk, an online platform where subjects can complete short experiments for small payments.  To increase data reliability, all subjects had a prior task approval rating of 90\%; they were approved for payment on at least nine out of ten tasks they had ever done previously on Mechanical Turk.  Each subject was presented with a target image at the center of their screen, followed by a 500 ms delay.  Subjects were then asked to click one of 7 response images (backgroundless canonical views of category exemplars) that matched the category of the target image.  After giving a response, subjects were shown a central fixation point for 500 ms before the next target image appeared.  Participating subjects were required to complete at least one block of trials, but were allowed to complete an unlimited number of additional blocks.  Each block of trials was presented with the same stimulus duration, which ranged between 25 ms and 2000 ms, and the number of trials within each block was chosen to keep the amount of time required for each block equal between stimulus duration conditions.  This had the effect of keeping the effective wage rate equal between conditions, which was approximately \$10/hour.  We additionally manually and algorithmically screened the data for cheating (for example, providing the same response to every target image, providing strings of identical responses, or failing to produce high entropy in the response distribution).  After screening the dataset consisted of 96,717 total responses.  For each block of trials we took the accuracy over the trials and excluded the first 5 responses.  The mean of the block accuracies for each presentation time are shown in Figure~\ref{fig:human_perf} and the error bars indicate the 95\% confidence interval of the standard error of the mean over block-accuracies.

\subsection*{Linear-SVM methodology}
For the linear support vector machine methodology (linear-SVM) we seek to measure the generalization performance of a linear classifier estimated on one subset of the data, the training set, and measured on another, the testing set.  By choosing a linear classifier the decision boundary is a simple linear function.  By imposing this simplicity on the classifier, effective representations will be those that allow this simple classifier to achieve high generalization performance, and performance is not confounded by the additional functional complexity introduced by a more complex decision function.  This procedure therefore requires the representation, rather than the decision function, to ``solve'' the problem.

In detail we train a linear support vector machine on 80\% of the images (1960 images total, 1568 training, 392 testing) using the scikit-learn (\url{http://scikit-learn.org}) wrapper to LIBLINEAR~\linebreak(\url{http://www.csie.ntu.edu.tw/~cjlin/liblinear}).  We used a squared hinge loss, a one versus rest procedure, a squared penalty on parameters, and selected the regularization parameter ($C$) by 3-fold cross-validation on the training set, and then re-estimating the linear-SVM using the optimal value for $C$ on the entire training set.  We then used the trained linear SVM to predict the categories of the testing images and report the mean classification accuracy over the testing set.  To measure statistical variation due to subsampling of image variation parameters we compute the testing set accuracy ten times, each time sampling 80\% of the images for training and the remaining for testing.  The ten samples are fixed for all representations and within each randomization we equalize the number of images from each category.

\begin{figure}[t]
\begin{center}
\includegraphics[width=.6\linewidth]{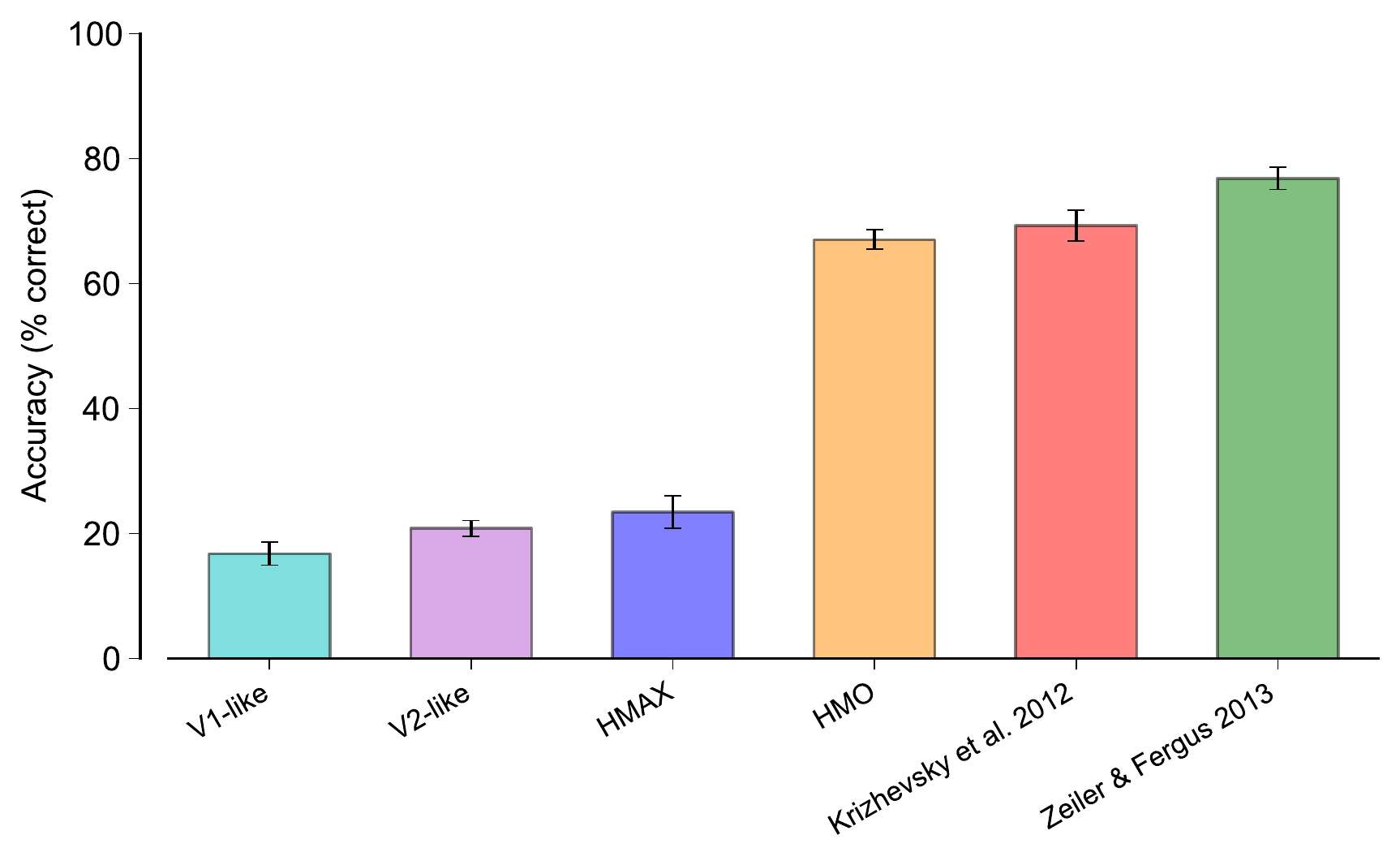}
\end{center}
\caption{{\bf Linear-SVM performance of model representations without sample or noise correction.}  Testing set classification accuracy averaged over 10 randomly-sampled test sets is plotted and error bars indicate standard deviation over the 10 random samples.  Chance performance is $\sim$14.3\%.  Unlike in Figure~\ref{fig:svm}, the model representations in this figure have not been modified to correct for sampling or noise.}
\label{fig:svm_absolute}
\end{figure}

\begin{figure}[h]
\begin{center}
\includegraphics[width=\linewidth]{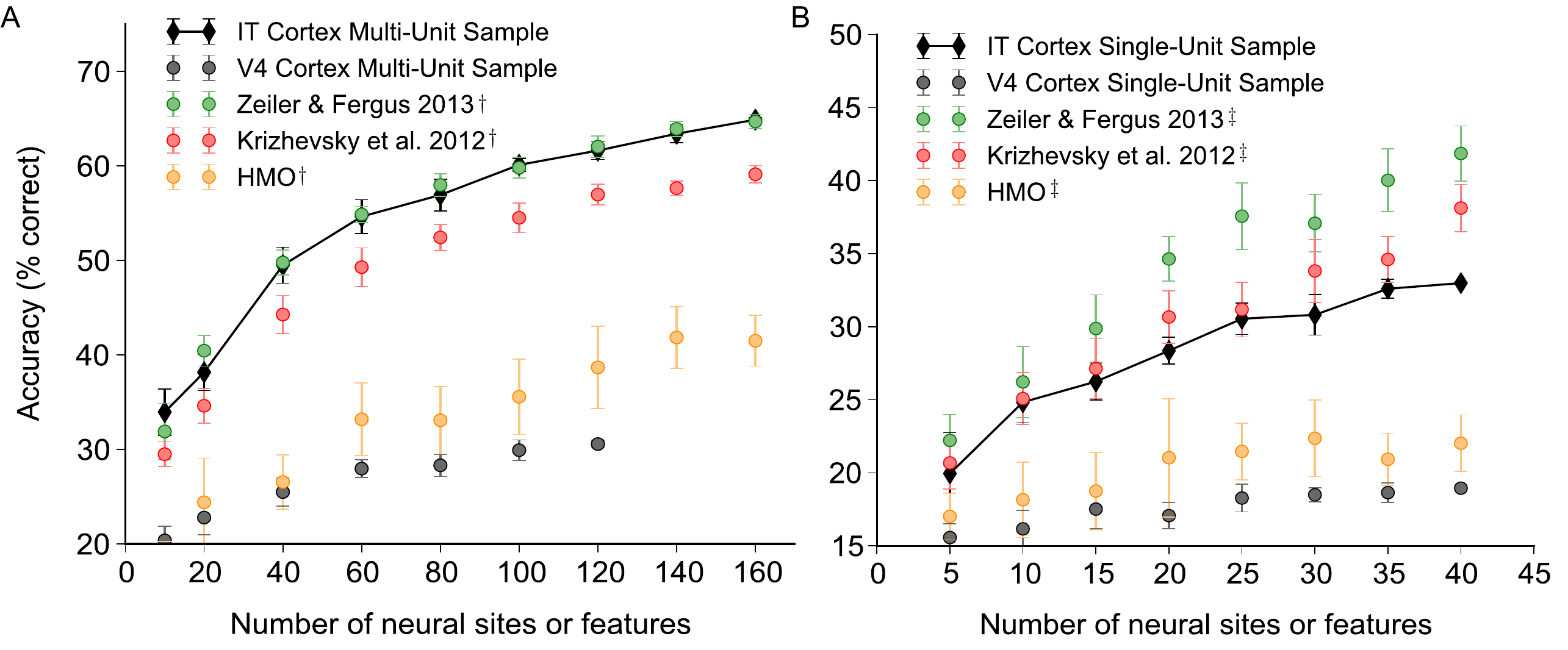}
\end{center}
\caption{{\bf Effect of sampling the neural and noise-corrected model representations for the linear-SVM analysis.}  We measure the mean testing-set linear-SVM generalization performance as we change the number of neural sites (for neural representations), or the number of features (for model representations).  Measured samples are indicated by filled symbols and measured standard deviations indicated by error bars.  Multi-unit analysis is shown in panel A and single-unit analysis in B.  The model representations are noise corrected by adding noise that is matched to the IT multi-unit measurements (A, as indicated by the $\dagger$ symbol) or single-unit measurements (B, as indicated by the $\ddagger$ symbol).  This analysis reveals a similar relationship to that found using the kernel analysis methodology (compare to Figure~\ref{fig:svm}).}
\label{fig:svm_subsampled}
\end{figure}

In Figure~\ref{fig:svm_absolute} we present the performance of the linear-SVM methodology on the model representations without correction for noise or trials.  In Figure~\ref{fig:svm_subsampled} we present the linear-SVM analysis as a function of sampling.  Figure~\ref{fig:svm_subsampled} is analogous to Figure~\ref{fig:subsampled}, but uses the linear-SVM methodology instead of the kernel analysis methodology.

\subsection*{Comparing IT multi-unit and single-unit representations}
\begin{figure*}[t]
\begin{center}
\includegraphics[width=.8\linewidth]{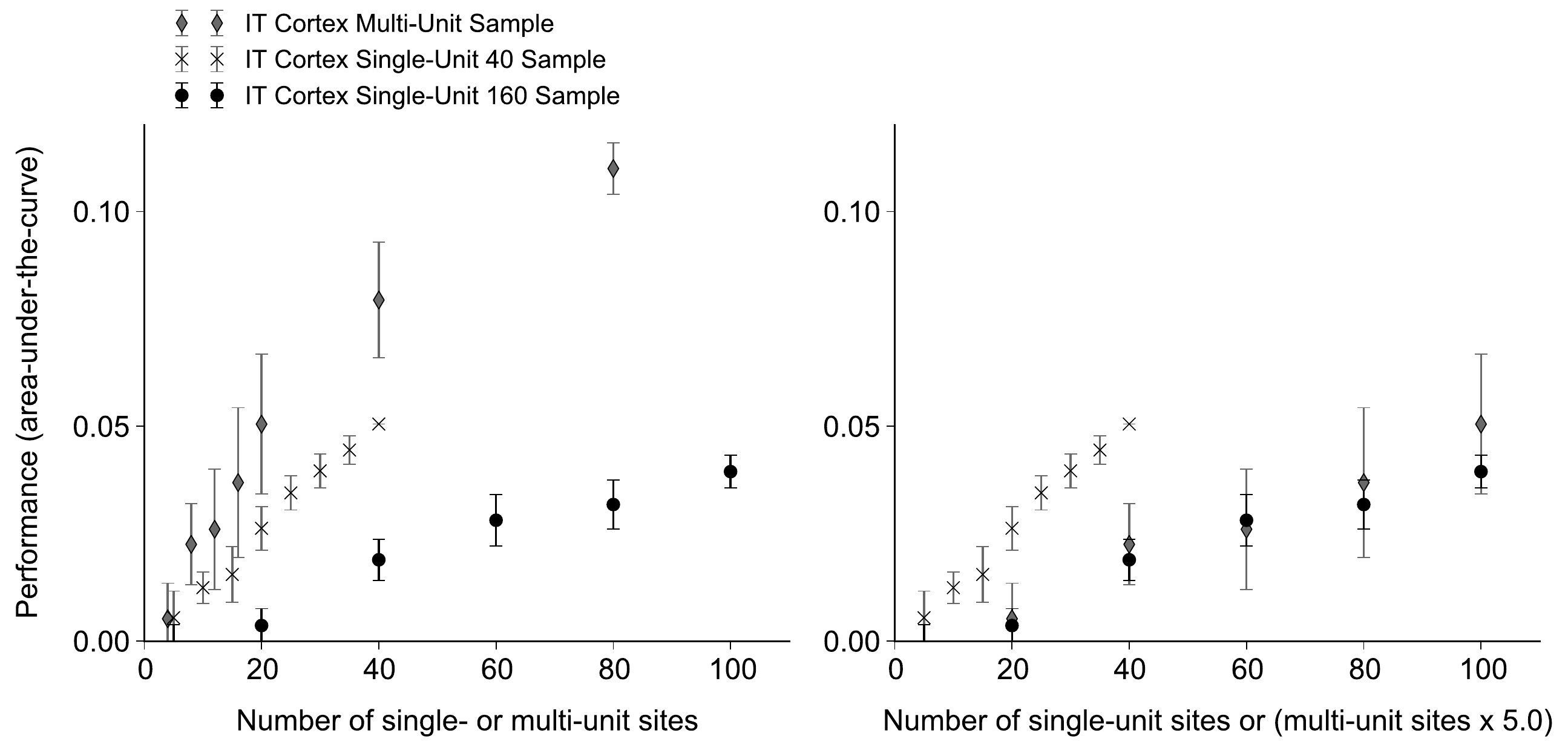}
\end{center}
\caption{{\bf Comparing IT multi-unit and single-unit representations}  In the left panel we plot the kernel analysis AUC as a function of the number of single- or multi-unit sites.  We plot results for two single-unit samples.  ``IT Cortex Single-Unit 160 Sample'' uses all 160 isolated single-units and ``IT Cortex Single-Unit 40 Sample'' uses the 40 most consistent (least noisy) single-units.  In the right panel we show the same data, but correct the number of multi-units to reflect an estimate of the number of single-units contributing to each multi-unit recording, thus plotting against the number of estimated neurons.  Unlike previous figures, these estimates have a fixed number of trials (6) for both single- and multi-unit samples.  Surprisingly, multi-unit recordings surpass single-unit recordings in performance (left) and are comparable in performance to unscreened single-units (right).}
\label{fig:multi_vs_single}
\end{figure*}

Here we directly compare the representational performance between the IT multi-unit sample and the IT single-unit sample.  Our previous analyses did not correct for the wide difference in trials between our multi- and single-unit samples, which have 47 and 6 trials, respectively.  In Figure~\ref{fig:multi_vs_single} we show the results of the kernel analysis area-under-the-curve measurements as a function of the number of single- or multi-unit sites and fixing each representation to 6 trials (responses are averaged over 6 randomly sampled trials).  For the single-unit sample we show the analysis using all 160 isolated single-units (``IT Cortex Single-Unit 160 Sample'') and the 40 most consistent (lowest noise) single-units (``IT Cortex Single-Unit 40 Sample'').  In the left panel of Figure~\ref{fig:multi_vs_single} we directly compare the number of multi-units to the number of single-units and show that the multi-unit representation outperforms the single-unit representation.  We can also introduce another correction between the multi-unit and single-units by attempting to compare directly the number of neurons that go into each measurement.  By using an independent dataset~\cite{Rust:2010uk} collected in our lab using single-electrode electrophysiology and comparing to our multi-electrode setup, we determine based on spike counts that each multi-unit in our sample is approximately 4-5 single-units.  Therefore, in the right panel in Figure~\ref{fig:multi_vs_single} we plot the results by multiplying the number of multi-units by 5.0.  This result indicates that the multi- and single-unit representations are roughly equivalent in performance when we do not screen single-units.  This remains surprising, as the physical averaging process (averaging multiple single-units in a multi-unit recording) produces a loss of information and may, a priori have been thought to reduce performance.  However, after this correction the screened single-units outperform the multi-units.

\subsection*{Processing time and energy consumption of computational models}
The processing time of DNNs as implemented on current hardware is comparable to that of the macaque and human visual systems.  The algorithm of Zeiler \& Fergus 2013, which is very similar in implementation to Krizhevsky et al. 2012, utilizes a high-performance GPU and processes batches of 128 images in 8.4 seconds, or 65 ms per image.  This processing time is shorter than the presentation times we use during the neural recordings and for our estimate of human performance (both 100 ms).  It is also shorter than the integration window we use to measure IT multi-unit responses, which we average between 70 ms and 170 ms post image onset.  Finally, it is also comparable to the latency of response in IT cortex to image presentation (typically 70 ms to 100 ms).  However, the 65 ms processing time for the DNN does not include the process of phototransduction (image capture) nor any communications latencies within the computer system (e.g. between system memory and GPU memory).  Behavioral response times of macaque to certain discrimination tasks are typically between 230 and 250 ms, and in humans are roughly 50 ms longer (see \cite{Thorpe:2001uv} and \cite{FabreThorpe:2011ho}).  It is likely that current DNN implementations could achieve response times less than those observed in macaque or human.

The energy efficiency of DNNs appears to be far worse than that of macaque or human.  Modern high-performance GPUs typically operate between 200 and 350 Watts (W) at load (100 to 125 W at idle).  Estimates of the power consumption of the entire human brain are typically around 20 W.  Using an estimate that the macaque ventral stream is around 1/40 the size of the human brain (the macaque brain is less than 1/10 the size of the adult human brain and a high estimate of the ventral stream size is 1/4 of the macaque brain), we estimate that macaque ventral stream operates around 0.5 W.  Assuming that the macaque brain operates at this power consumption rate during visual behavior, we estimate that the macaque ventral stream is around 400 times or between 2 and 3 orders of magnitude more energy efficient than current DNN implementations (however, see \cite{Farabet:2011fk} for progress on energy efficient DNN implementations).

\pagebreak
\bibliography{plos}
\pagebreak
\end{document}